\documentclass[reprint,aps,prl]{revtex4-2}
\usepackage{natbib}
\usepackage{bm}
\usepackage{tensor}
\usepackage{amsthm}
\usepackage{amsmath}
\usepackage{amssymb}
\usepackage{amsfonts}
\usepackage{mathrsfs}
\usepackage{mathtools}
\usepackage{graphicx}
\DeclareMathAlphabet{\mathpzc}{OT1}{pzc}{m}{it}

\DeclareMathOperator{\tr}{\text{tr}}

\newcommand{\set}[1]{\left\{#1\right\}}

\newcommand{\bmm}[1]{\bm{\mathrm{#1}}}

\begin{document}
\title{Indexed singular value bounds on scattering operators:\\ How many channels can a photonic device support?}
\author{Paul Virally} 
\author{Sean Molesky}
\affiliation{Department of Engineering Physics, Polytechnique Montr\'{e}al, Montr\'{e}al, Qu\'{e}bec H3T 1J4, CAN}
\author{Pengning Chao}
\affiliation{Department of Mathematics, Massachusetts Institute of Technology, Cambridge, Massachusetts 02139, USA} 
\author{Alessio Amaolo}
\affiliation{Department of Chemistry, Princeton University, Princeton, New Jersey 08544, USA}
\author{Alejandro W. Rodriguez}
\affiliation{Department of Electrical and Computer Engineering, Princeton University, Princeton, New Jersey 08544, USA}
\email{sean.molesky@polymtl.ca} 
\begin{abstract}
\noindent
Spectral properties of scattering operators, and their dependence on geometry, are of crucial importance to photonic design, enabling low-rank approximations and improved understanding of achievable power and information transfer. 
Here, we develop a method to bound indexed singular values (channel amplitudes) of the Green operator, $W$-operator, and proposed $P$-operator, for arbitrarily structured linear media. 
The approach yields computable upper bounds on the $n^{th}$ singular value, for any given $n$, that capture the complexity of multi-channel tradeoffs and competing scattering effects.
As illustrations of the framework, channel bounds are provided for multi-wavelength three-dimensional ``mediating'' volumes (up to $64\,\lambda^3$, mimicking communication waveguide-like and metasurface-like configurations), power transfer between $9\,\lambda^3$ source and receiver volumes, and applied to elucidate the performance of a planewave angle discrimination problem (bounding the smallest singular value, or condition number, of a fixed input space). 
In addition to these exemplary uses, the approach is directly applicable to bounds on information theoretic objectives such as Shannon capacity and Fisher information, as well as computational guarantees, such as error limits for reduced-order models. 
\end{abstract}
\maketitle
\noindent
Motivations for wanting photonic devices capable of controlling as many distinct inputs as possible are simple and compelling: e.g., (1) data transfer, which is in large part a multiplexing design problem, is both a direct precursor of computation, and an area where photonic solutions are likely to play an increasingly important role in the near future\;\cite{kuznetsov2024roadmap,shekhar2024roadmapping,hai2025next,onodera2026multimode}; (2) at a high level, many photonic devices are communication systems---mapping a set of input waves to a set of output waves---and communication performance is strongly determined by the ability of a system to sort and transform various inputs\;\cite{miller2012all,wetzstein2020inference,sharma2025universal}; (3) advances realized in machine learning over the past decade have opened previously unimaginable possibilities in signal processing and inference\;\cite{karanov2018end,arya2024end,chi2025neural}, removing many former operational constraints for encoding via field characteristics. 
\\ \\
With such motivations in mind, the abstraction of channels, or more precisely singular value decomposition (SVD), provides a natural language for linear system design. 
Namely, the fact that SVD expresses a linear operator as a countable set of  decoupled one-dimensional transformations tends to simplify both mental and phenomenological descriptions\;\cite{suryadharma2017singular,chen2018generalizing,zhang2018singular}.
More concretely, many electromagnetic processes are concisely stated in terms of the Green operator\,\footnote{The designation of Green operator, as opposed to the usual Green function, is intended to distinguish between the linear operator, which includes integration, and the kernel, which does not. 
The Green function, or more widely Green's function, is conventionally used to denote both objects.}---the inverse of the Maxwell operator---allowing questions concerning phenomenological understanding to be reframed as questions of why the operator possesses certain characteristics. 
Via SVD, such descriptions are broken down and simplified into the behaviour of orthogonal ``channels''---mappings between specific (orthogonal) input currents in some sending volume and specific (orthogonal) output fields in some receiving volume---that can be analysed independently.
This set of channels is, moreover, automatically ordered in terms of quantitative importance by the corresponding singular values or ``channel amplitudes'' (Fig.\,1): the first $n$ components of a SVD sum to give the most accurate rank $n$ approximation that can be made under the nuclear or Frobenius (Hilbert-Schmidt, or trace) norm, establishing a ``count'' of distinct wave transformations. 
\\ \\
Due to the effective equivalence between the Frobenius norm and common optimization objectives like overlap functions, power transfer, and energy density, SVD thinking is innately suggested in a variety of application areas\;\cite{angeris2021heuristic,gertler2025many}, and seems almost inescapable when considering problems involving network capacity\;\cite{telatar1999capacity,tang2010mimo}, dimensionality reduction\;\cite{martinsson2011randomized,jaradat2021tutorial,marzban2025inverse}, and forecasting and recommendation\;\cite{sun2018analysis,lahmiri2018minute,yuan2019singular}. 
For similar underlying reasons, many approaches to large-scale simulation are also  efficiently described through the lens of creating light-weight approximations to linear operators that are faithful to within a controllable level of precision, making notions rank and pseudo-rank highly useful\;\cite{chew1993nepal,martinsson2007fast,liu2021butterfly,xue2023fullwave,mavrikakis2025surface}.  
\\ \\
Beyond the connection of these overarching concepts to photonics, channel counts, and accompanying amplitude metrics, equally serve as a common currency for net-transfer (basis independent) phenomena and a range of input-output device design problems.
Conceptually, almost any process that does not depend on specific field realizations typically admits a trace formulation\;\cite{kruger2012trace,yao2022trace}, which can be restated as a functional sum over the singular values of the scattering operator.
As such, channel amplitudes provide an intriguing point of comparison for a diverse collection of physics, such as emission phenomena encompassing fluorescence, thermal radiation, heat transfer, and Casimir forces\;\cite{hamam2007coupled,rodriguez2012fluctuating,strekha2022trace}, as well as scattering phenomena encompassing cloaking, extinction, absorption\;\cite{molesky2020t,hanisch2022relative,strekha2024limitations} (Fig.\,1, \emph{Formulation}). 
In particular, under an additive noise model (detector dominated noise), the number of channels above threshold sharply dictates Shannon capacity\;\cite{ehrenborg2020physical}, and is also closely related to mutual information\;\cite{pinkard2025information} and Fisher information (see \emph{Applications}). 
Practically, for input to unconstrained output type applications---transform some known set of inputs into a corresponding set of distinguishable, but otherwise arbitrary, outputs---a range of objectives hinge on realizing a mediator (e.g. meta-optic, coupler, splitter, etc.) to improve transmission into a fixed receiver: e.g. to concentrate power within a volume across some preset range of angles (lensing)\;\cite{lin2021computational}, or to route distinguishable modes entering on a common waveguide into spatially separated regions (spatial multiplexing)\;\cite{guo2025advances}. 
So long as the form of the fields produced within the receiver is inconsequential, the performance of photonic devices for this sort of design problem reduces to achievable channel amplitudes (see \emph{Formulation}). 
Accordingly, there are a number of scenarios related to communication\;\cite{franceschetti2009capacity,jiang2011singular,akrout2022achievable,ehrenborg2021capacity,gustafsson2025shadow}, imaging\;\cite{o2002information,kabuli2026designing} and classification\;\cite{eltaieb2020efficient} wherein engineering channel amplitudes is of substantial interest. 
\begin{figure*}[htp!]
    \includegraphics[width=1.0\linewidth]{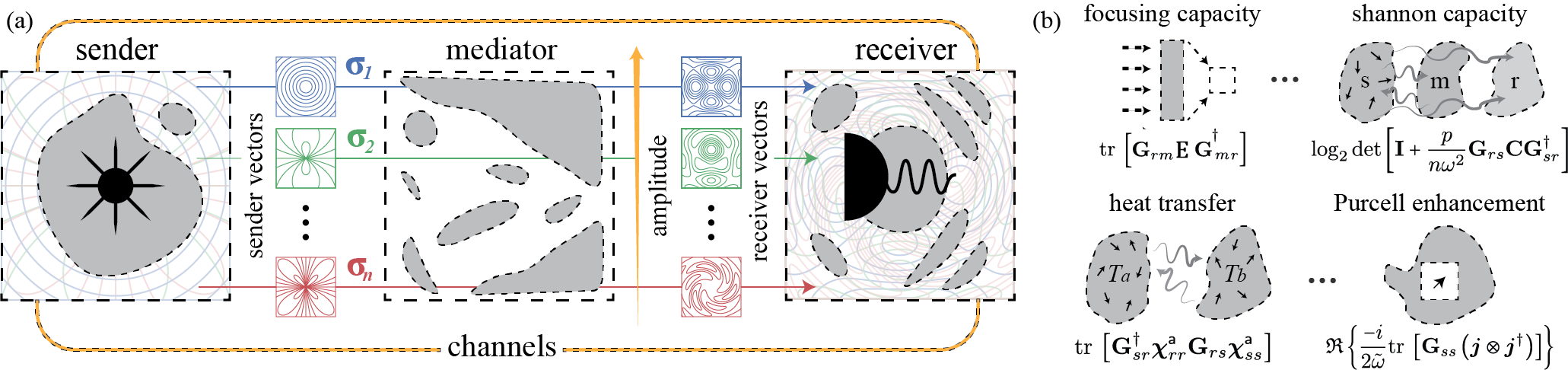}
    \caption{
    \label{fig:articleSchematic} 
    \textbf{Channels and applications.} 
    Panel (a) illustrates the meaning of a channel in the sender-mediator-receiver convention used throughout the article. 
    Each channel consists of a pair of connected input-output fields (one in the sender and one in the receiver), that are orthogonal to the fields of every other channel within their respective domains of definition, along with a channel amplitude (the singular value $\sigma_{k}$) that dictates the strength of the mapping. 
    Working in terms of channels leads to a fully decoupled description of linear operation. 
    Panel (b) depicts four phenomena, among many, that are efficiently described as sums over singular values: heat transfer\;\cite{molesky2020fundamental}, Purcell enhancement\;\cite{van2012dyadic}, lensing\;\cite{bertero1982resolution}, and Shannon capacity\;\cite{amaolo2024maximum}. 
    A description of the notation used in the figure is provided at the beginning of the \emph{Formulation} section.
    }
\end{figure*} 
\\ \\
Granting the utility of SVD based perspectives in photonic design, several questions are brought forward. 
Principally, to what extent can the channels of a device be engineered\;\cite{poon2005degrees,miller2013complicated,miller2013establishing,miller2019waves,molesky2021comm,seyedinnavadeh2024determining,zhao2024channel,guo2025unitary}?
Or, more simply, how many channels can a device (of some bounded size and composed of some given materials) possibly support above a specified cutoff\;\cite{miller2000communicating,ehrenborg2017fundamental,ehrenborg2021capacity,amaolo2024maximum,kuang2025bounds}? 
The present article advances ongoing efforts to understand the design landscape of photonic systems along this second line.
Specifically, we provide the first protocol to generally answer the following question that begins to capture the tradeoffs imposed by the need to handle multiple inputs and antagonistic scattering phenomena. 
\emph{Given volumes of space that may contain arbitrarily structured materials, how large can the $n^{th}$ singular value of a scattering operator be?}
\\ \\
The findings presented below show that by combining the Courant-Fischer-Weyl min-max principle with convex relaxation bounds\;\cite{chao2022physical,miller2026fundamental}, quantitative limits on the $n^{th}$ singular value of any common photonic scattering operator are computable for moderately sized photonic devices (here up to $64\,\lambda^{3}$).  
After briefly sketching the main ideas behind the approach, and mentioning some notable relations and links to prior work, five explanatory applications are explored. 
\\ \\
(a) To begin we show that, as a simple consequence of a ``rank removal'' corollary of the Courant-Fischer-Weyl (CFW) min-max principle, previously derived bounds on integrated near-field heat transfer can be upgraded to indexed (power transfer) channel bounds\;\cite{molesky2020fundamental}. 
The resulting analytic forms provide instructive approximations to more exact CFW  programs that we anticipate will be pursued in future works.
In connection with these results, we also propose the definition of a new ``field power'' scattering operator ($\bmm{P}_{rs}$, \emph{Formulation: Power Scattering}), and prove the channel amplitudes of this operator are bounded by unity in any electromagnetic system (including loss-less idealizations).
\\ \\
(b) Bounds on the number of channels having amplitude greater or equal to a specified value are given for long design regions inserted between thin, unstructured, sender and receiver volumes, investigating the extent to which a material similar to silicon can possibly enhance transmission over free space.
As may be guessed, the results are dramatic: when the ``waveguide'' is present, a separation of $5\lambda$ compares favourably with free space transmission down to distances smaller than $\lambda /4$.
\\ \\
(c) Channel amplitude bounds are investigated for metasurface-like configurations of wide-thin design regions inserted between wide-thin (unstructured) sender and receiver pairs with varying aspect ratios and separations. 
We observe that while the largest channel amplitudes remain relatively stable, increasing the volume of the ``metasurface'' can greatly increase the number of usable channels (qualified as singular value bounds larger than $0.1$). 
For small separations this effect saturates with increasing device area, hinting that while larger ``non-local'' cell volumes are likely a key to realizing greater functionality (c.f. Refs.\;\cite{kwon2018nonlocal,overvig2020multifunctional,chen2025nonlocal}) there is a point of diminishing returns. 
\\ \\
(d) The fourth, more applied, example examines a two dimensional sensing problem mimicking laser awareness\;\cite{li2023transcending}: a system composed of a thin receiver and interposed mediator (i.e. a metaoptic and a sensor) is freely optimized to distinguish between elements of a specified set of incoming plane waves spanning a $30^{\circ}$ cone of incidence. 
Once again, device size is seen to have an important effect. 
As the width of the mediator increases from $2\,\lambda$ to $16\,\lambda$, the maximal number of identifiable angles (the limit number of channels having amplitude greater than $0.1$) grows from $5$ to $17$. 
\\ \\
(e) Finally, connecting to fundamental limits of near-field heat transfer\;\cite{molesky2020fundamental,venkataram2020fundamental}, per-channel bounds are computed for power transfer---the indexed components of the $\Phi\left(\omega\right)$ flux factor that appears in radiative heat transfer---between arbitrarily structured cubic source and receiver volumes separated by a fixed vacuum gap.
While prior results are known to be nearly tight for thin-films at frequencies near a high quality surface plasmon-polariton resonances, power transfer between finite three-dimensional objects has yet to be investigated with contemporary convex relaxation machinery~\cite{chao2022physical}. 
Comparing against the flux factor between a pair of wavelength radius silica balls at $20\,\mu m$, corresponding to a weak surface plasmon-polariton resonance, computed bounds are found to be between one and two orders of magnitude larger depending on separation, suggesting that a larger number of channels may be activated by optimized geometries. 
Near-field asymptotics, for the regimes we are able to access, show a dependence of roughly $\propto d^{-5/4}$ as a function of the separation distance $d$ (part way between the $\propto d^{-1}$ separation dependence of balls\;\cite{narayanaswamy2008thermal}, and the $\propto d^{-2}$ separation dependence of flat half-spaces\;\cite{polder1971theory}). 
For high-quality materials, the bounds also allow the possibility that a small, fixed, number of channels can maintain approximately unit amplitudes at very large separation distances (up to $1000\,\lambda$). 
\\ \\
Throughout these examples, either by direct comparison with inverse design or indirect comparison to known results, computed indexed channel amplitude bounds are found to be highly predictive of achievable performance, hinting at potential synergies with large-scale multi-functional photonic device design\;\cite{gertler2025many,dalklint2025potential,chao2025blueprints}.
\\ \\
Additional details concerning these examples, program formulations, technical arguments, and computational procedures are provided in \emph{Supporting Material}.
\section{\S Formulation}
\label{sec:Forumulation}
\noindent
\emph{Formulation} begins by presenting our notation and the central technical results on which this work is based. 
Following this overview, the $P$-operator $\bmm{P}_{rs}$ is proposed as a common basis for comparing channel amplitudes of photonic scattering operators. 
\subsection{Formulation: Notation and Conventions}
\noindent
Throughout the text, vector quantities (e.g. an incident electric field $\bm{e} = \bm{E}\left(\bm{r},\omega\right)$ or initial current density $\bm{j} = \bm{J}\left(\bm{r},\omega\right)$ in Fig.\;\ref{fig:articleSchematic}) are denoted by uncapitalized bold and italicized letters. 
Vector spaces and subspaces are shown by italicized capital letters. 
Linear operators between vector spaces (e.g. the $W$-operator mapping between initial and total polarization current densities) are denoted by capitalized bold letters, e.g. $\bmm{W}$. 
Daggers on operators stand for adjoints, and daggers on vectors stand for partner linear functionals (i.e. Dirac bras under the implicitly assumed inner product). 
$\lVert\bm{y}\rVert_{2}$ is the $L^{2}$-norm of the vector $\bm{y}$. 
An $\mathsf{a}$ superscript is used for the anti-symmetric part of an operator $\bmm{A}^{\mathsf{a}} = \left(\bmm{A}-\bmm{A}^{\dagger}\right)/2i$, typically the positive definite (or semi-definite) part under our Fourier transform convention. 
$\bmm{A}^{/2}$, for a positive semi-definite operator $\bmm{A}$, is defined by the relation $\left(\bmm{A}^{/2}\right)^{2} = \bmm{A}$.
If $\bmm{A} = \bmm{V}\bm{\Sigma}\bmm{U}^{\dagger}$ is the singular value decomposition of a scattering operator $\bmm{A}$, then the columns of the unitary operator $\bmm{U}$ are the orthogonal source vectors, the columns of the unitary operator $\bmm{V}$ are the orthogonal receiver vectors, the ordered terms of decomposition expression 
$$
    \bmm{A} = \bmm{V}\bmm{\Sigma}\bmm{U}^{\dagger} = \sum_{k}\sigma_{k}\,\bm{v}_{k}\otimes\bm{u}^{\dagger}_{k},
$$
are the system channels, and the elements of the diagonal operator $\Sigma$---the diagonalization of $\bmm{A}$ in the $\left(\bmm{U},\bmm{V}\right)$ basis---$\left\{\sigma_{k}\right\}$ are the channel amplitudes.
As an overloaded notation, $\sigma_{n}\left(\bmm{A}\right)$ is also used for the function extracting the $n^{th}$ singular value of the operator $\bmm{A}$. 
$j$ and $k$ subscripts are also occasionally used to denote summation following the Einstein contraction convention.
Although essentially all of the presented theory remains valid in either the temporal or frequency domain, all examples and computations are carried in the frequency domain, and the entire study was approached with a frequency domain mindset. 
\\ \\
$\bmm{G}^{\circ}$ is the vacuum (or free space) Green operator\,
\footnote{If so desired, $\bmm{G}^{\circ}$ can be taken to be any background Green function if $\bm{\chi}$ is defined as the scattering material not included in the background. 
This sort of redefinition can be helpful in problems where a partition of an overall system is fixed. 
However, for the optimization programs defined later in this supplement, it is typically easier to take $\bmm{G}^{\circ}$ to be vacuum and assert that the variable $\chi$ is in fact known in certain volumes.}---the complete linear mapping, including integration, that inverts the Maxwell equations, supposing radiative boundary conditions, up a multiplicative factor of $iZ/k_{\circ}$. 
That is, we take the convention of defining the wave operator as the mapping on the electric field $\bm{e}$ that appears on the left of the relation  
$$
    \left(\frac{\nabla\times\nabla\times}{k_{\circ}^{2}} - \bmm{1}\right)\bm{e} = \frac{iZ}{k_{\circ}}\bm{j},
$$
the electromagnetic wave equation for $e^{-i\omega t}$ harmonics with both sides divided by $k_{\circ}^{2} = \left(2\pi/\lambda\right)^{2}$\,\footnote{The primary motivation for this convention is that the wavelength $\lambda$ becomes a natural length scale for numerical discretization.}, and $\bmm{G}^{\circ}$ as inverse of the wave operator---$\bm{e} = \left(iZ/k_{\circ}\right)\bmm{G}^{\circ}\bm{j}$ with $Z$ the impedance of free space. 
Operating under the assumption of linear response, which we make throughout, $\bmm{\chi}$ denotes the linear susceptibility, mapping electromagnetic fields to polarization currents via the relation $\bm{j} = -i\left(k_{\circ}/Z\right)\bm{\chi}\,\bm{f}_{t}$. 
E.g., $\bm{j} = -i\left(k_{\circ}/Z\right)\bm{\chi}^{ee}\bm{e} = -i\left(k_{\circ}/Z\right)\int_{\bmm{r}'}\chi_{jk}^{ee}\left(\bm{r},\bm{r}',\omega\right)e_{k}\left(\bm{r}'\right) = \left(1-\epsilon\left(\omega\right)\right)e_{k}\left(\bm{r}\right) = j_{k}\left(\bm{r}\right)$ in the case of an isotropic, non-magnetic, medium.
$\bmm{G}$ is used for the Green operator in the presence of some particular (although perhaps undetermined) device (material configuration). 
\\ \\
In reference to the sender-mediator-receiver description of Fig.\;\ref{fig:articleSchematic}, $r,\,m$ and $s$ subscripts denote that the domain (source vector space, right subscript of a pair) or codomain (target vector space, left subscript of a pair) is restricted to the receiver, mediator, or source respectively. 
When only one space of an operator is restricted, a $u$ subscript (universal) is added for clarity. 
$\bmm{1}$ is used for the identity operator (e.g. $\bmm{1}_{rr}$ is the identity restriction to the volume of the receiver).
\\ \\
Basic scattering theory language is used freely. 
Readers unfamiliar with this terminology, or who find our usage odd, are referred to Ref.\;\cite{molesky2020t}, noting that we now use $\bm{\chi}$ for electromagnetic susceptibility quantities instead of $\mathbb{V}$, $\bmm{1}$ for the identity instead of $\mathbb{I}$, and $\bmm{W}$ for the $W$-operator instead of $\mathbb{T}\mathbb{V}^{-1}$. 
\\ \\ 
Filling out the as of yet undefined quantities of Fig.\;\ref{fig:articleSchematic}, $\tilde{\omega}$ is a complex frequency (which allows the Purcell enhancement formula to be applied to a Lorentzian window function\;\cite{chao2023maximum}), $\otimes$ is the outer product (i.e. the matrix formed by a column and a row), $n$ and $p$ are the noise and signal power in reference to the current-current covariance matrix $\bmm{C}$\;\cite{amaolo2024maximum}, and $\bmm{E} = \sum_{k}\bm{e}_{k}\otimes\bm{e}^{\dagger}_{k}$ is the covariance matrix of electric fields incident on the lensing region.
Suppressing unit conversions and multiplicative factors that are independent of material structuring, the equations given in Fig.\;\ref{fig:articleSchematic} yield the following physical quantities.
The first expression is the integrated intensity (focusing) of a given set of electric fields into a receiving volume of space; the second expression is the Shannon capacity of a photonic system; the third expression is the radiative heat transfer between a source object and receiver object held at distinct temperatures in vacuum; the fourth expression is the Purcell enhancement for an initial source distribution $\bm{j}$.
\subsection{Formulation: Power Scattering}
\noindent
Tying into the forthcoming discussion of \emph{Prior Methods}, whenever the sender and receiver are well defined, channel amplitudes are reported by 
\begin{equation}
    \bmm{P}_{rs} = 2\,\bm{\chi}_{rr}^{\mathsf{a}/2}\,\bmm{G}_{rs}\,\bm{\chi}_{ss}^{\mathsf{a}/2},
    \label{eq:untOpt}
\end{equation}
which we refer to as the $P$-operator or field power operator.
\\ \\
Our preference for $\bmm{P}_{rs}$ over other scattering operators is based on two considerations. 
The first amounts to using a scale natural to relevant physics. 
Via the fluctuation-dissipation theorem, the spectral radiative heat-transfer between a sender at temperature $T_{s}$ and reviver at temperatures $T_{r}$ is 
$$
    h\left(\omega\right) = \Theta\left(\omega, T_{r}, T_{s}\right)\Phi_{rs}\left(\omega\right),
$$ 
where $\Theta\left(\omega, T_{r}, T_{s}\right)$ is the difference in the expected spectral thermal energy of source and receiver states,
$$
    \Theta\left(\omega, T_{r}, T_{s}\right) = \hbar\omega\left(n_{be}\left(\omega,T_{r}\right) - n_{be}\left(\omega,T_{s}\right)\right)/\left(2\pi\right)
$$ 
with $n_{be}\left(\omega,T\right) = \left(\exp\left(\hbar\omega/\left(k_{B}T\right)\right)-1\right)^{-1}$ the Bose-Einstein expectation number, and $\Phi_{rs}\left(\omega\right)$ is a unitless flux factor encoding the response characteristics of the photonic system\;\cite{polder1971theory,rytov1988principles2,pendry1999radiative}. 
Using Eq.\;\eqref{eq:untOpt},
$$
    \Phi_{rs}\left(\omega\right) = \tr\left(\bmm{G}^{\dagger}_{rs}\bm{\chi}^{\mathsf{a}}_{rr}\bmm{G}_{rs}\bm{\chi}^{\mathsf{a}}_{ss}\right) = \tr\left(\bmm{P}^{\dagger}_{sr}\bmm{P}_{rs}\right) 
    .
$$
Because $h\left(\omega\right)$ defines a spectral power transfer, and $\Theta\left(\omega, T_{r}, T_{s}\right)$ covers all units and thermal characteristics, it follows that $\bmm{P}_{rs}$ is a unitless operator describing power scattering (as it relates to a field values). 
\\ \\
In more detail, $\Phi_{rs}\left(\omega\right)$ decomposes into two terms: the power absorbed by the receiver for any unitary basis of polarization currents in source,
$$
    \tr\left(\bmm{G}^{\dagger}_{rs}\bm{\chi}^{\mathsf{a}}_{rr}\bmm{G}_{rs}\right) 
    =\!\int_{V_{r}}\!\!d\bm{r}\,
    E^{*}_{j}\left(\bm{r},\omega\right)\chi_{jk}^{\mathsf{a}}\left(\bm{r},\omega\right)E_{k}\left(\bm{r},\omega\right),
$$
and an overall scaling factor, which in the case of a source consisting of a single isotropic material becomes $k_{\circ}\,\Im\left(\chi_{s}\right)/Z$---the real part of the conductivity $\sigma_{s}$ of the source material.
For a fixed input power $p$, Ohm's law indicates that in the absence of geometric structuring the intensity of currents created in the source is $\sigma_{s}p = j^{2}$, with $j^{2}$ denoting the current intensity. 
As such, this scaling factor can be understood as a conversion between input power and  initial current intensity in the source, which the field absorption factor then translates into the total power transferred to the receiver. 
That is, when viewed as process, $\Phi_{rs}\left(\omega\right)$ takes an initial drive power existing in the source and maps it to a total power delivered in the receiver.
By the relation of $\bmm{P}_{rs}$ to $\Phi_{rs}\left(\omega\right)$, $\bmm{P}_{rs}$ therefore functions as a unitless ``power scattering'' operator for fields. 
\\ \\
Connecting with recent literature, the definition of $\bmm{P}_{rs}$ given by Eq.\;\eqref{eq:untOpt} is also found to be well suited to photonic Shannon capacity. 
In Ref.\;\cite{amaolo2024maximum}, the spectral Shannon capacity of a photonic system was reported in terms of $\bmm{G}_{rs}$.
This choices raises two irritations.
First, using the notation of Ref.\;\cite{amaolo2024maximum}, to evaluate the $\bmm{G}_{rs}$-capacity expression with unitful operators, the noise level $N$ must have units of the squared electric field, while the covariance constraint value $C$ must be unitless---$C$ is not a power level as a unit conversion is required to place $C$ and $N$ on an equal footing. 
Because $\bmm{P}_{rs}$ is unitless, this asymmetry does not arise for a $\bmm{P}_{rs}$ channel matrix. 
Following Refs.\;\cite{telatar1999capacity,amaolo2024maximum}, assuming Gaussian noise with diagonal and uniform covariance of power normalized magnitude $n$, the $\bmm{P}_{rs}$-capacity of a photonic system is bounded by the optimization 
\begin{align}
    \text{sc}\left[\bmm{P}_{rs}\right]\leq&\max_{\bmm{D}\succeq 0}\,\log_{2}\det\left[\bmm{1}_{rr} + \frac{1}{n}\,\bmm{P}_{rs}\bmm{D}\bmm{P}_{sr}^{\dagger}\right]
    \label{eq:ShannonCap} \\
    &~\text{s.t.}\, \tr\left[\bmm{P}_{us}\bmm{D}\bmm{P}_{su}^{\dagger}\right] \leq p,
    \nonumber
\end{align}
where $\bmm{D}$ is the ``drive'' covariance of initial (input) source currents converted into power units\,\footnote{Note that $\bmm{D}$ is not the power drawn by the system. 
The actual power, drawn from some external mechanism, is determined by the total currents, which include all scattering phenomena, and not the initial currents represented by $\bmm{D}$. 
That is, prior to actually specifying the system geometry, the amount of power that will be drawn ($\tr\left[\bmm{P}_{us}\bmm{D}\bmm{P}_{su}^{\dagger}\right]$) is not fully determined. 
Nevertheless, because the singular values of $\bmm{P}_{rs}$ are bounded by unity, the power transferred to receiver is always smaller than the power required to support the (hypothetical) distribution of initial currents, making $\tr\left[\bmm{D}\right] \leq 1$ a useful constraint.}, $\bmm{P}_{rs}\bmm{D}\bmm{P}_{sr}^{\dagger}$ is the covariance matrix of power transferred to the receiver, $\bmm{P}_{us}\bmm{D}\bmm{P}_{su}^{\dagger}$ is the covariance matrix of total transferred power---power transferred to the receiver, absorbed by the source material, or lost to the mediating environment, and $p$ is the total power drawn by the system.  
Second, in the limit of a lossless material there is presently no method to bound the singular values of $\bmm{G}_{rs}$ that allows for arbitrary material structuring (over a vanishing bandwidth\;\cite{shim2019fundamental,chao2023maximum}): because there is no strict relation between the singular values of $\bmm{G}_{rs}$ and consumed power, $\bmm{G}_{rs}$-capacity bounds become arbitrarily large as this idealization is approached.   
$\bmm{P}_{rs}$-capacity does not have this issue. 
Restating Eq.\;\eqref{eq:ShannonCap} in terms of singular values 
\begin{align}
    &\text{sc}\left[\bmm{P}_{rs}\right]\leq\max_{\bm{d}}\,\sum_{k=1}^{\infty}\log_{2}\left(1 + \frac{d_{k}}{n}\,\sigma_{k}^{2}\left(\bmm{P}_{rs}\right)\right)
    \nonumber \\
    &~\text{s.t.}\,\left(\forall k\right)\,d_{k}\geq 0~\&~\sum_{k=1}^{\infty}d_{k}\left(\sigma_{k}^{2}\left(\bmm{P}_{rs}\right) + \alpha_{k}^{2}\left(\bmm{P}_{rs}\right)\right)\leq p,
    \nonumber
\end{align}
with $d_{k}$ the $kk$ expansion coefficient of $\bmm{D}$ in the right singular basis of $\bmm{P}_{rs}$, and $\alpha_{k}^{2}\left(\bmm{P}_{rs}\right)$ the (non-negative) power transferred by a $k^{th}$ right singular vector of $\bmm{P}_{rs}$ to anything other than the receiver\,\footnote{That is, $(\sigma_{k}^{2}\left(\bmm{P}_{rs}\right) + \alpha_{k}^{2}\left(\bmm{P}_{rs}\right)$ is the total transferred power for the $k^{th}$ right singular vector of $\bmm{P}_{rs}$.}.
Defining $x_{k} =d_{k}\left(\sigma_{k}^{2}\left(\bmm{P}_{rs}\right) + \alpha_{k}^{2}\left(\bmm{P}_{rs}\right)\right)$, this optimization becomes
\begin{align}
   \text{sc}\left[\bmm{P}_{rs}\right]\leq &\max_{\bm{x}}\,\sum_{k=1}^{\infty}\log_{2}\left(1 + \frac{x_{k}}{n}\frac{\sigma_{k}^{2}\left(\bmm{P}_{rs}\right)}{\sigma_{k}^{2}\left(\bmm{P}_{rs}\right) + \alpha_{k}^{2}\left(\bmm{P}_{rs}\right)}\right)
    \nonumber \\
    &~\text{s.t.}\,\left(\forall k\right)\,x_{k}\geq 0~\&~\sum_{k=1}^{\infty}x_{k}\leq p,
    \nonumber
\end{align}
which is efficiently solved by the water-filling\;\cite{floudas2013deterministic} given bounds on $\sigma_{k}^{2}\left(\bmm{P}_{rs}\right)/\left(\sigma_{k}^{2}\left(\bmm{P}_{rs}\right) + \alpha_{k}^{2}\left(\bmm{P}_{rs}\right)\right)$.
More simply, as a consequence of the algebraic form, it follows that $\sigma_{k}^{2}\left(\bmm{P}_{rs}\right)/\left(\sigma_{k}^{2}\left(\bmm{P}_{rs}\right) + \alpha_{k}^{2}\left(\bmm{P}_{rs}\right)\right) \leq 1$.
Therefore,
\begin{equation}
    \text{sc}\left[\bmm{P}_{rs}\right]\leq p/\left(n\,\!\ln\left(2\right)\right):
    \label{eq:simpleShannonBound}
\end{equation}    
The spectral $\bmm{P}_{rs}$-capacity is always limited by the ratio of the available power to the noise power. 
\\ \\ 
The second consideration is mostly practical.
Generalizing the trace bounds of Ref.\;\cite{molesky2020fundamental} via the rank removal corollary stated below, the positivity of scattered power proves that the algebraic conclusions encountered in the Shannon capacity context hold in general: $\sigma_{n}\left(\bmm{P}_{rs}\right)\leq 1$ for any photonic system. 
In addition to formalizing the intuition that material structuring can at best result in unit efficiency power transfer, this bound is useful for comparing between system configuration choices;
the channel amplitudes of $\bmm{P}_{rs}$ always lie in the unit interval. 
\\ \\
Conversion between channel amplitudes in terms of $\bmm{P}_{rs}$, and channel amplitudes in terms of $\bmm{G}_{rs}$ and $\bmm{W}_{rs}$, for isotropic media, are accomplished by the formulas
\begin{align}
    &\sigma_{n}\left(\bmm{G}_{rs}\right) = \,\sigma_{n}\left(\bmm{P}_{rs}\right)/ \sqrt{4\,\Im\left(\chi_{s}\right)\Im\left(\chi_{r}\right)}
    \label{eq:channelAmplitudeConversions} \\
    &\sigma_{n}\left(\bmm{W}_{rs}\right) = \sigma_{n}\left(\bmm{P}_{rs}\right)\left|\chi_{r}\right|/\sqrt{4\,\Im\left(\chi_{s}\right)\Im\left(\chi_{r}\right)}.
    \nonumber
\end{align}
In cases where either the source or receiver object is undefined---e.g. we are given an incident field, or for computational simplicity we suppose currents densities in vacuum---we revert to reporting channel amplitudes in terms of either $\bmm{G}_{rs}$ or $\bmm{W}_{rs}$ as appropriate. 
\subsection{Formulation: Courant-Fischer-Weyl Min-Max Principle}
\noindent
The underpinning of our proposed framework is the Courant-Fischer-Weyl (CFW) min-max principle, and an associated ``rank-removal'' corollary, which combine to restate the $n^{th}$ singular value of a linear operator as a min-max or max-min optimization.  
Intuitively, the first $n$ right singular vectors of a compact linear operator $\bmm{A}:X\rightarrow Y$ form an $n$-dimensional subspace, $U^{n}$. 
Take $Q^{n-1}$ to be any $(n-1)$-dimensional subspace of $X$ and $Q^{n-1}_{\perp}$ to be its orthogonal complement.
Because $X = Q^{n-1}\oplus Q^{n-1}_{\perp}$, some linear combination of elements of $U^{n}$ must reside in $Q^{n-1}_{\perp}$---$Q^{n-1}\cap U^{n}$ is at most an $\left(n-1\right)$-dimensional subspace---and so a direction orthogonal to $Q^{n-1}\cap U^{n}$ within $U^{n}$ must always exists. 
Let $\bm{x}$ be a the unit vector along this direction. 
As each component of $\bm{x}$ will be scaled by a factor of at least $\sigma_{n}$, $\max_{\bm{y}\in Y}\Re\left(\bm{y}^{\dagger}\bmm{A}\bm{x}\right)$, with $\lVert\bm{y}\rVert_{2} = 1$, is $\geq\sigma_{n}$.
\\ \\
\emph{CFW Min-Max Principle}---Let $\bmm{A}:X\rightarrow Y$ be a compact linear operator between complex Hilbert spaces with ordered singular values $\sigma_{1}\left(\bmm{A}\right) \geq \sigma_{2}\left(\bmm{A}\right) \geq \cdots $. 
$\bmm{A}^\dagger \bmm{A}$ is then a compact self-adjoint operator with ordered eigenvalues $\sigma_{1}^{2}\left(\bmm{A}\right)\geq\sigma_{2}^{2}\left(\bmm{A}\right) \geq \cdots$. 
The $n^{th}$ eigenvalue of $\bmm{A}^\dagger \bmm{A}$ is equal to the following min-max (max-min) optimization over the $n$-dimensional subspace $Q^{n}\subseteq X$ and its orthogonal complement $Q^{n}_{\perp}$\;\cite{teschl2014mathematical}:
\begin{subequations}
    \begin{align}
        \sigma_{n}^{2}\left(\bmm{A}\right) &= \min_{Q^{n-1}\subseteq X}\max_{~\bm{x}\in Q^{n-1}_{\perp}}\bm{x}^\dagger\bmm{A}^\dagger\bmm{A}\bm{x}
        \label{eq:CFWminmax}\\
        &= \max_{Q^{n}\subseteq X }\min_{\bm{x}\in Q^{n}}\bm{x}^\dagger\bmm{A}^\dagger\bmm{A}\bm{x}
        \label{eq:CFWmaxmin},
    \end{align}
\end{subequations}
with $\lVert\bm{x}\rVert_{2} = 1$.
Stripping away the outer optimization gives an upper (resp. lower) bound on $\sigma_{n}^{2}$:
\begin{equation}
    \min_{\bm{x}\in Q^{n}}\bm{x}^\dagger\bmm{A}^\dagger\bmm{A}\bm{x}\leq\sigma_{n}^{2}\left(\bmm{A}\right)\leq\max_{\bm{x}\in Q^{n-1}_{\perp}}\bm{x}^\dagger\bmm{A}^\dagger\bmm{A}\bm{x}.
    \label{eq:CFWeigbounds}
\end{equation}
Equivalent bounds, as verified by the Cauchy-Schwarz inequality, can also be stated in terms of $\bmm{A}$ with an additional optimization over $\bmm{y}\in Y$:
\begin{equation}
     \min_{\bm{x}\in Q^{n}}\max_{\bm{y}\in Y}~\bm{y}^{\dagger}\bmm{A}\bm{x}\leq\sigma_{n}\left(\bmm{A}\right)\leq\max_{\bm{x}\in Q^{n-1}_{\perp}}\max_{\bm{y}\in Y}~\bm{y}^\dagger\bmm{A}\bm{x}.
     \label{eq:CFWsvbounds}
\end{equation} 
\noindent
\emph{Rank Removal}---Let $\bmm{A}:X\rightarrow Y$ be a linear operator between complex Hilbert spaces possessing a singular value decomposition.
If the rank of $\bmm{B}:X\rightarrow Y$ is $\leq n$,  
\begin{equation}
    \sigma_{n+1}\left(\bmm{A}\right)\leq\sigma_{1}\left(\bmm{A}-\bmm{B}\right).
    \label{eq:rankRemove}
\end{equation}
\begin{proof}[Proof of Rank Removal.]
    Let $\bmm{A} = \sum_{i}\gamma_{i}\,\bm{v}_{i}\otimes\bm{u}_{i}^{\dagger}$ be the singular value decomposition of $\bmm{A}$, and $\bmm{B} = \sum_{k=1}^{n}\kappa_{k}\,\bm{a}_{k}\otimes\bm{b}_{k}^{\dagger}$ be the singular value decomposition of $\bmm{B}$. 
    Take $Q^{n}$ to be the subspace of $X$ spanned by $\left\{\bm{b}_{k}\right\}$, and $Q^{n}_{\perp}$ to be its orthogonal complement.
    By the second version of the CFW min-max principle, 
    \begin{align}
        \sigma_{n+1}\left(\bmm{A}\right)&\leq\max_{\bm{x}\in Q^{n}_{\perp}}\,\max_{\bm{y}\in Y}\,\bm{y}^{\dagger}\bmm{A}\bm{x} = 
        \max_{\bm{x}\in Q^{n}_{\perp}}\,\max_{\bm{y}\in Y}\,\bm{y}^{\dagger}\left(\bmm{A}-\bmm{B}\right)\bm{x}
        \nonumber \\
        &\leq\sigma_{1}\left(\bmm{A}-\bmm{B}\right),
        \nonumber
    \end{align}
    with the middle equality following from $\bmm{B} = \bmm{0}$ in $Q^{n}_{\perp}$.
\end{proof}
\noindent
In what follows, the CFW min-max principle is combined with rank removal to derive bounds on the $n^{th}$ channel amplitude of $\bmm{G}_{rs}$ based on the $T$-operator expansion 
\begin{equation}
    \bmm{G}_{rs} = \bmm{G}^{\circ}_{rs} + \bmm{G}^{\circ}_{rm}\bmm{T}_{mm}\bmm{G}^{\circ}_{ms},
    \label{eq:gtExp}
\end{equation}
and $\bmm{P}_{rs}$ (or $\bmm{W}_{rs}$) by the nested-scattering expansion
\begin{equation}
    \bmm{W}_{rs} = \bmm{W}^{(i)}_{rr}\bmm{\chi}_{rr}\bmm{G}^{\circ}_{rs}\bmm{W}_{ss}
\end{equation}
where $\bmm{W}^{(i)}_{rr}$ is the $W$-operator of the receiver in the absence of the source and mediator, Eq.\;\eqref{eq:WInvertedA}. 
Depending on the application in question, this is done by either asserting $Q^{n-1}$ to be formed by some collection of the singular vector basis of $\bmm{G}^{\circ}_{ms}$ and then applying a convex relaxation over a quadratically constrained quadratic program describing the possible response characteristics of $\bmm{T}_{mm}$ (c.f. Supporting Material), by first performing these steps and then freely optimizing over $Q^{n-1}$, or by using rank removal to iteratively remove the largest channels of some known intermediate operator.   
\subsection{Formulation: Field Sources and Restricted Outputs}
\noindent
As sketched in the laser awareness example treated in \emph{Applications}, the CFW approach is equally applicable to settings wherein there is no source region, but rather only incident waves, as well as design problems wherein detection is restricted to a particular subspace (e.g. the detector only registers certain field patterns).
In the first instance, this is done by factoring Eq.~\eqref{eq:gtExp} as $\left(\bmm{1}_{rr} + \bmm{G}^{\circ}_{rm}\bmm{T}_{mm}\bmm{1}_{mm}\right)\bmm{G}^{\circ}_{us}$ and then replacing the $\bmm{G}^{\circ}_{us}$ factor appearing on the right with the matrix formed by concatenating the input (source) fields incident on the device,  $\bmm{S} = \left[\bm{s}_{1},\ldots,\bm{s}_{n}\right]$.
Taking the additional step of supposing a fixed receiver, the transfer operator then becomes $\bmm{S}_{ru} + \bmm{G}^{\bullet}_{rm}\bmm{T}_{mm}\bmm{S}_{mu}$ where $\bmm{G}^{\bullet}_{rm}$ is the Green operator for the receiver in the absence of the mediator\;\cite{sun2025fast}, $\bmm{S}_{ru}$ is the set of fields created in the receiver in the absence of the mediator, and $\bmm{S}_{rm}$ are the incident fields appearing in the design region of the mediator.
Restrictions on detection, e.g. limiting detectable output to a certain subspace of polarization current densities, may be implemented in a similar manner by the inclusion of subspace projector $\bmm{O}$ on the operator output, $\bmm{O}\left(\bmm{S}_{ru}+\bmm{G}^{\circ}_{rm}\bmm{T}_{mm}\bmm{S}_{mu}\right)$. 
\\ \\
Once the appropriate transfer operator has been formulated, e.g. the matrix of the realized outputs associated with each source $\bmm{A} = \bmm{S}_{ru} + \bmm{G}^{\bullet}_{rm}\bmm{T}_{mm}\bmm{S}_{mu}$, several aspects of the (multiplexing) design problem can be analysed in terms of its channel amplitudes. 
Maximizing $\tr\left(\bmm{A}\right) = \sum_{k=1}^{n}\sigma_{k}\left(\bmm{A}\right)$, the sum of the singular values of $\bmm{A}$, or $\tr\left(\bmm{A}^{\dagger}\bmm{A}\right) = \sum_{k=1}^{n}\sigma_{k}^{2}\left(\bmm{A}\right)$, the Frobenius norm, increases the total transmission over the vector space spanned by the columns of $\bmm{S}$ as quantified by either the field amplitudes (sum) or the field intensities (squared sum). 
Maximizing $\sigma_{n}\left(\bmm{A}\right)$, the smallest singular value, is equivalent to the max-min objective of increasing the output of the unit-norm input combination in the span of $\bmm{S}$ most susceptible to additive noise. 
Removing $\bmm{S}$, $\bmm{G}_{rm}^{\circ}\bmm{T}_{mm}$ describes the related problem of transmission into the receiving region given complete freedom over the input fields\,\footnote{There are of course many other objectives that could also be plausibly be considered in relation to such problems. 
For example, one might be interested in considering the total error for some exact set of inputs (instead of the linear span), or favouring certain types of output fields without requiring exact agreement. 
The former objective can be written in singular value form by considering $n$ rank-one operators.
The latter objective can be treated by adding weightings on top of a projection operator into the receiver region.}\,\footnote{
For emission like applications, such as antennas\;\cite{capek2017minimization} or controlling interaction with optically addressable qubits\;\cite{chakravarthi2020inverse}, relevant transfer operators typically place greater emphasis on the source, either by making it part of the designable part of the device, or by considering it as both a sending and receiving component (self-transfer).  
Once this is done, a common aim is to make the trace as large as possible. 
(E.g. maximizing the local density of states at a given location is equivalent to maximizing the vector trace of the co-located anti-symmetric part of the Green operator\;\cite{chao2023maximum}.)}.
\\ \\
Certain end-to-end type applications, comprising e.g. the design of on-chip data links\;\cite{nossek2013chip} and ``inseparable'' electromagnetic process like near-field heat transfer\;\cite{jin2017overcoming}, can similarly be interpreted as communication devices wherein all components are to some extent designable\,\footnote{In most scenarios we have in mind the true freedom must invariably be constrained in some way to produce a tractable optimization formulation.}. 
Again, depending on the extent to which encoding and decoding is possible (or required), there is generally interest in either transferring as much power as possible over a large number of inputs (i.e. maximizing the Frobenius norm\;\cite{kruger2012trace,yao2022trace} as done in \emph{Applications}), or optimizing some capacity metric\;\cite{ehrenborg2021capacity}.
For this second class of objective, the relative distribution of channel amplitudes is of critical importance, and being able to tractably predict how physics limits what distributions are possible was a major impetus for the development of the method proposed here. 
\\ \\
Through these and similar interpretations and operator compositions, it is possible to handle a broad class of cases of practical interest for all major scattering operators. 
However, there is as of yet no clear prescription for satisfactorily treating a receiver that can only measure intensities---the root issue being that the mapping from fields to intensities is non-linear. 
More technically, although relations do exist between singular values and Hadamard products, we are not currently aware of an argument that correctly captures the intuition that neglecting phase information must cause certain distinct communication channels to merge that is tractable with the current CFW optimization strategy.  
While it is fairly obvious that phase sensitive detection is ultimately superior to intensity only detection, and that consequently any derived limits for measurement sensitivity or information theoretic quantities likely remain valid, it is less clear what the consequences are at the level of individual channel bounds. 
Developing this understanding is an evident direction for future work. 
\section{\S Prior Methods}
\noindent
We are aware of only two prior quantitative methods for bounding indexed singular values of electromagnetic scattering operators allowing for arbitrarily structured materials: ``vacuum separation'' bounds and two-object interaction bounds via the Frobenius norm. 
Although both of these techniques remain valuable in terms of conceptual implications and ease of application, sample calculations have found the CFW strategy to be consistently tighter, often by orders of magnitude.
In the case of vacuum separation bounds, this follows mathematically. 
The $1\,|\,n$ inequality (see below) from which these limits follow can be derived by combining a particular choice of $Q^{n-1}$ with a further relaxation of the Courant-Fischer-Weyl (CFW) min-max principle. 
The final form is essentially equivalent to assuming that every channel utilizes a (unique) perfectly resonant response. 
In the case of Frobenius norm bounds, the distinct nature of the approximations used makes proving a general ordering more difficult. 
However, the square-root decay characteristic of this approach is unphysically slow based on comparison with either vacuum separation bounds or CFW. 
For large $n$, Frobenius norm will vastly overestimate achievable channel amplitudes. 
\subsection{Prior Methods: Vacuum Separation Bounds}
\noindent
In Ref.\;\cite{kuang2025bounds}, shape-independent upper bounds on $\sigma_{n}$ for current-current ($W$-operator) communication between two volumes (that can be separated by a spherical surface) are derived based on the following \emph{$1\,|\,n$ inequality}. 
\\ \\
\emph{$1\,|\,n$ Inequality}---Let $\bmm{A}:X\rightarrow Y$ and $\bmm{B}:Y\rightarrow Z$ be linear operators between complex vector spaces.  
\begin{equation}
    \sigma_{n}\left(\bmm{AB}\right)\leq\min\left\{\sigma_{1}\left(\bmm{A}\right)\sigma_{n}\left(\bmm{B}\right),\sigma_{n}\left(\bmm{A}\right)\sigma_{1}\left(\bmm{B}\right)\right\}.
    \label{eq:1nInequailty}
\end{equation}
\begin{proof}[Proof of $1\,|\,n$ Inequality.]
    Select the first $n-1$ right singular vectors of $\bmm{B}$ to form $Q^{n-1}$. 
    By CFW, 
    \begin{align}
        \sigma_{n}^{2}\left(\bmm{AB}\right)&
        \leq\max_{\bmm{x}\in Q^{n-1}_{\perp}}\bmm{x}^{\dagger}\bmm{B}^{\dagger}\bmm{A}^{\dagger}\bmm{A}\bmm{B}\bmm{x}
        \nonumber \\
        &\leq
        \sigma_{1}^{2}\left(\bmm{A}\right)
        \max_{\bmm{x}\in Q^{n-1}_{\perp}}\bmm{x}^{\dagger}\bmm{B}^{\dagger}\bmm{B}\bmm{x}
        = \sigma_{1}^{2}\left(\bmm{A}\right)\sigma_{n}^{2}\left(\bmm{B}\right).
        \nonumber
    \end{align}
    The second inequality forming the minimum follows by replacing $\bmm{AB}$ with $\bmm{B}^{\dagger}\bmm{A}^{\dagger}$.
\end{proof}
\noindent
Next, taking $s$ and $r$ subscripts to denote restrictions into the sender and receiver volumes, consider the formal description of the $W$-operator of the system in block form:
\begin{equation}
    \bmm{W} = 
    \begin{bmatrix}
        \bmm{1}_{rr}-\bmm{\chi}_{rr}\bmm{G}^{\circ}_{rr} & -\bmm{\chi}_{rr}\bmm{G}^{\circ}_{rs} \\
        -\bmm{\chi}_{ss}\bmm{G}^{\circ}_{sr} & \bmm{1}_{ss}-\bmm{\chi}_{ss}\bmm{G}^{\circ}_{ss} 
    \end{bmatrix}^{-1}.
    \nonumber
\end{equation}
Using block matrix inversion, 
\begin{equation}
    \bmm{W} = 
    \begin{bmatrix}
        \bmm{W}_{rr} & \bmm{W}_{rr}^{(i)}\bmm{\chi}_{rr}\bmm{G}^{\circ}_{rs}\bmm{W}_{ss} \\
        \bmm{W}^{(i)}_{ss}\bmm{\chi}_{ss}\bmm{G}^{\circ}_{sr}\bmm{W}_{rr} & \bmm{W}_{ss}
    \end{bmatrix}, 
    \label{eq:WInvertedA}
\end{equation}
where $\bmm{W}^{(i)}$ is the $W$-operator for the appropriate isolated object (e.g., $\bmm{W}^{(i)}_{ss}$ is the $W$-operator for the source in the absence of any receiver), and $\bmm{W}$ is the dressed operator accounting for the presence of both scattering volumes---$\bmm{W}_{rs}$ maps initial currents in the source volume to self-consistent (total) polarization currents in the receiver volume, accounting for all scattering effects. 
\\ \\
The bounds of Ref.\;\cite{kuang2025bounds} combine these two results. 
Applying the $1\,|\,n$ inequality twice to the source-receiver block of Eq.\;\eqref{eq:WInvertedA}, 
\begin{equation}
    \sigma_{n}\left(\bmm{W}_{rs}\right)\leq
    \sigma_{1}\left(\bmm{W}_{rr}^{(i)}\bmm{\chi}_{rr}\right)\sigma_{n}\left(\bmm{G}^{\circ}_{rs}\right)\sigma_{1}\left(\bmm{W}_{ss}\right).
    \label{eq:sepBounds}
\end{equation}
Because both $\sigma_{1}\left(\bmm{W}_{rr}^{(i)}\bmm{\chi}_{rr}\right)$ and $\sigma_{1}\left(\bmm{W}_{ss}\right)$ are bounded for any physical material, e.g. Ref.\;\cite{miller2014fundamental} and Ref.\;\cite{molesky2020fundamental}, Eq.\;\eqref{eq:sepBounds} proves the crucial point that communication is always limited by free space propagation across the vacuum separation. 
No matter how the sender and receiver are designed, channels amplitudes are ultimately restricted by the free space separation to 
\begin{align}
    &\sigma_{n}\left(\bmm{W}_{rs}\right) \leq \zeta_{r}\zeta_{s}\,\sigma_{n}\left(\bmm{G}^{\circ}_{rs}\right)/\left|\chi_{s}\right|, 
    \label{eq:1NinqResults} \\
    &\sigma_{n}\left(\bmm{P}_{rs}\right) \leq 2\sqrt{\zeta_{r}\zeta_{s}}\,\sigma_{n}\left(\bmm{G}^{\circ}_{rs}\right),
    \nonumber
\end{align}
where $\zeta_{s} = \left|\chi_{s}\right|^{2}/\Im\chi_{s}$ and $\zeta_{r} = \left|\chi_{r}\right|^{2}/\Im\chi_{r}$ are the maximal material response factors\;\cite{miller2014fundamental}. 
\\ \\
However, for wavelength scale design volumes, which constitutes the most meaningful application of Eq.\;\eqref{eq:sepBounds} in light of the singular value growth characteristics elucidated in Ref.\;\cite{miller2025tunnelling}, Eq.\;\eqref{eq:sepBounds} will almost certainly be loose. 
First, as shown below, the necessity of coupling power from the source to the receiver is incompatible with the assumption that both volumes optimally amplify each left and right singular vector: $\sigma_{1}\left(\bmm{W}_{rr}^{(i)}\right)$ and $\sigma_{1}\left(\bmm{W}_{ss}\right)$ are not independent, so that generally enhancement proportional to $\sqrt{\zeta_{r}\zeta_{s}}$ in the $P$-operator is not possible (see \emph{Applications: Rank Removal Upgrade}). 
Second, there are a range of design problems that are best served by structuring an intermediate mediator volume (which is neither part of the source or the receiver). 
Lumping a mediator into one side of the two volume description of Eq.\;\eqref{eq:sepBounds} (e.g. as a super-source) will typically lead to a over-estimation of the singular values of $\bmm{G}^{\circ}_{rs}$---the implied design volume will be too large, and the separation will be too small. 
Finally, the blanket assumption of optimally resonant response for $\bmm{W}_{rr}^{(i)}$ and $\bmm{W}_{ss}$ obviates all difficulties associated handling many distinct input fields.
For large devices, this relaxation may well be quite accurate (e.g., near perfect absorbers do exist\;\cite{molesky2019bounds}). 
For wavelength scale communication systems it is likely very strong\;\cite{molesky2021comm}.
\\ \\
With all this said, the understanding provided Eq.\;\eqref{eq:sepBounds} should not be discounted; the characteristics of $\bmm{G}^{\circ}_{rs}$ between the various design volumes plays a critical role in determining the sort of functionality that can be engineered, regardless of how the materials are structured\;\cite{pendry1999radiative}. 
Using the CFW formulation, consideration of Eq.\;\eqref{eq:WInvertedA} does indeed appear to lead to reasonably predictive and computational tractable channel amplitude bounds for $\bmm{P}_{rs}$.
\subsection{Prior Methods: Two-Object Interaction Bounds via the Frobenius Norm}
\noindent
Somewhat trivially, bounds on the $n^{th}$ singular value of a linear operator are also set by bounds on the square of its $n^{th}$ partial Frobenius norm: if $c_{1}\geq c_{2}\geq\ldots\geq c_{n}\geq 0$, and $\sum_{k=1}^{n}c_{k} = v$, then the largest $c_{n}$ can be is $v/n$. 
Since the squared singular values of the partial Frobenius norm of any linear operator $\bmm{A}$ respect this condition,
\begin{equation}
    \sigma_{n}\left(\bmm{A}\right)\leq\lVert\bmm{A}\rVert_{\mathtt{F}_{n}}/\sqrt{n}.
    \label{eq:FrobeniusBounds}
\end{equation}
Consequently, Eq.\;\eqref{eq:FrobeniusBounds} can be used to derive bounds on the $n^{th}$ singular value of an electromagnetic transfer operator given a bound on its associated partial Frobenius norm (c.f.\;\cite{miller2016fundamental,molesky2020fundamental}). 
\\ \\
The main approach used to determine bounds on the partial Frobenius norm of $\bmm{P}_{rs}$ in past literature is a combination of ``basis-chaining'', modelled on the von Neumann trace inequality provided in the Supporting Material, and the common structure shared by power transfer between two objects and the constraint that scattered power must always be positive restricted to a single component of a multi-component system. 
\\ \\
Paraphrasing the argument of Ref.\;\cite{molesky2020fundamental}, the spectral flux factor associated with heat transfer between two objects is
$$
    \Phi\left(\omega\right) = 4\tr\left(\bmm{G}^{\dagger}_{sr}\bmm{\chi}_{rr}^{\mathsf{a}}\bmm{G}_{rs}\bmm{\chi}_{ss}^{\mathsf{a}}\right) = \tr\left(\bmm{P}^{\dagger}_{sr}\bmm{P}_{rs}\right) = \lVert\bmm{P}_{rs}\rVert_{\mathtt{F}}^{2}.
$$
Using the scattering equalities $\bmm{\chi}_{rr}\bmm{G}_{rs} = \bmm{W}_{rs}$ and $\bmm{W}_{rr}\bmm{\chi}_{rr} = \bmm{T}_{rr}$, and the sender-receiver block of Eq.\;\eqref{eq:WInvertedA}, the above may be rewritten as 
$$
    \Phi\left(\omega\right) = 4\tr\left(\bmm{T}_{ss}^{\dagger}\bmm{G}^{\circ\dagger}_{sr}\bmm{T}_{rr}^{(i)\dagger}\bmm{T}_{rr}^{(i)}\bmm{G}^{\circ}_{rs}\bmm{T}_{ss}\right) / \left(\zeta_{s}\zeta_{r}\right), 
$$
where $\bmm{T}_{rr}^{(i)}$ is the $T$-operator of the receiver in the absence of the sender---$\bmm{T}_{ss}$ depends on the receiver, but $\bmm{T}_{rr}^{(i)}$ is independent of the sender.
\\ \\
Because $\Phi\left(\omega\right)$ describes the power absorbed by the receiver object for thermally generated (stochastic and incoherent) currents originating in the sender, this same form arises in the constraint that the power scattered from a incoming field incident on the sender in the presence of the receiver must be positive. 
Again using block multiplication and defining
$$
    \bmm{X}_{ss} = \bmm{G}^{\circ\dagger}_{sr}\bmm{T}_{rr}^{(i)\dagger}\bmm{T}_{rr}^{(i)}\bmm{G}^{\circ}_{rs}/\zeta_{r},
$$
isolation of the sender-sender block of the positivity of scattered power constraint implies that 
$$
    \bmm{T}_{ss}^{\mathsf{a}} - \bmm{T}_{ss}^{\dagger}\bmm{T}_{ss}/\zeta_{s} - \bmm{T}_{ss}^{\dagger}\bmm{X}_{ss}\bmm{T}_{ss}\succeq 0, 
$$
where the final term is $\Phi\left(\omega\right)$ scaled by $\zeta_{s}$. 
In Ref.\;\cite{molesky2020fundamental}, an argument relying on the basis changing and this scattering inequality is used to conclude that
\begin{equation}
    \lVert\bmm{P}_{rs}\rVert_{\mathtt{F}_{n}}^{2}\leq
    \sum_{k=1}^{n}
    \begin{cases}
        \,~~~~~~~~1 & \kappa_{k} \geq 1 \\
        \,4\kappa_{k}/\left(1 + \kappa_{k}\right)^{2} & \kappa_{k}\leq 1
    \end{cases},
    \label{eq:oldBoundsFrobenius}
\end{equation}
where $\kappa_{k} = \zeta_{s}\zeta_{r}\,\sigma_{k}^{2}\left(\bmm{G}^{\circ}_{rs}\right)$. 
As shown in the next section, the Frobenius norm proves to be unnecessary, and these bounds apply individually to each channel. 
\section{\S Applications}
\label{sec:applications}
\begin{figure*}[htp!]
    \begin{center}
        \includegraphics[width=\linewidth]{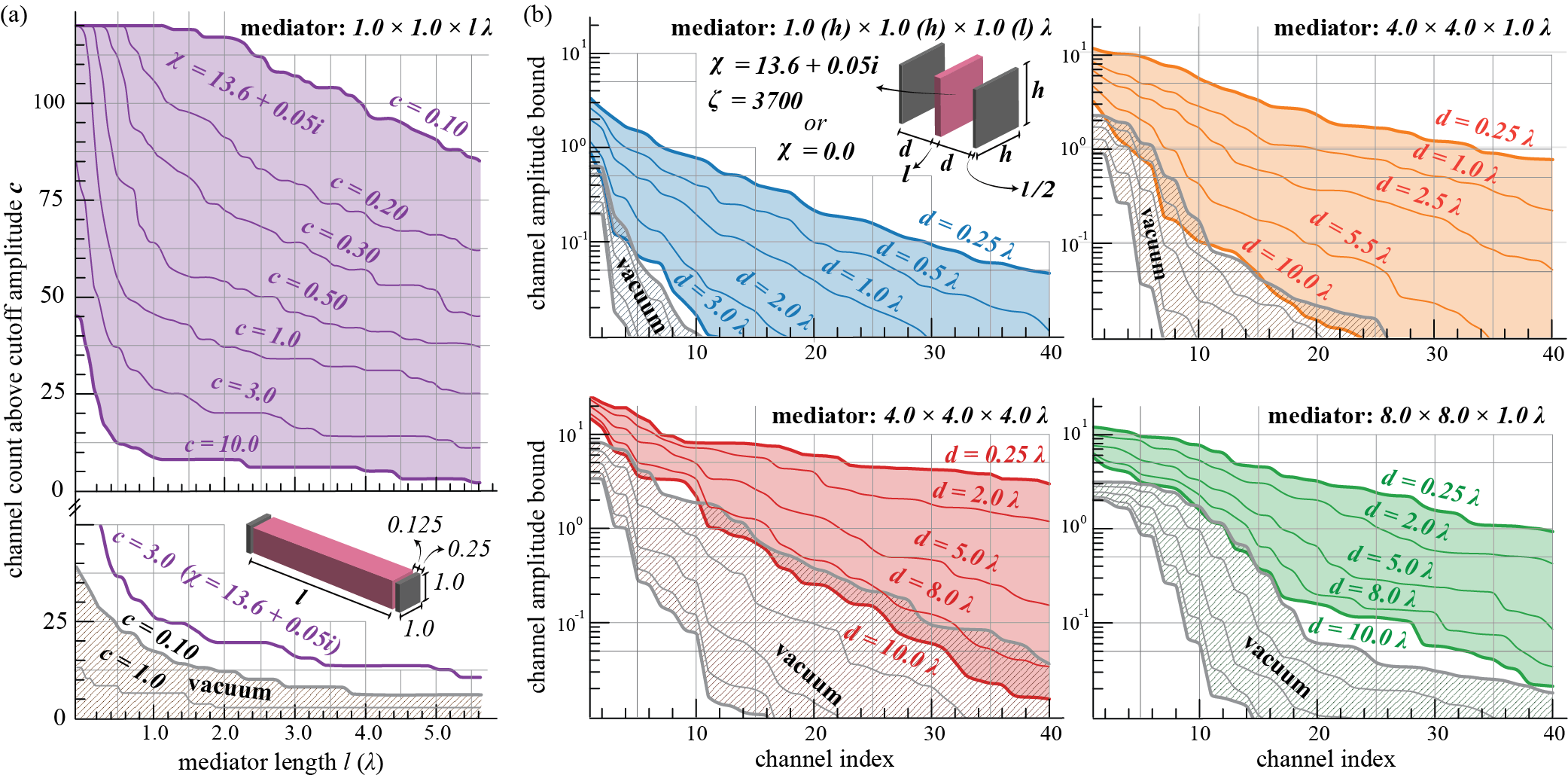}
        \caption{\label{fig:svdBounds}
        \textbf{Limit $\bmm{G}_{rs}$ channel amplitudes for common three-dimensional design scenarios.} 
        (a) 
        Upper bounds on channel numbers for arbitrarily structured waveguide-like (top: $\chi = 13.6 + 0.05i$, silicon-like, bottom: $\chi = 0$, vacuum) mediator volumes for varying lengths, $l$, and cutoff amplitudes, $c$. 
        The mediator (schematic in the bottom panel) is always separated from the unstructured sender and receiver regions (on either side) by $\lambda/4$ . 
        The purple $c = 3.0$ count line in the bottom panel is simply recopied from the top panel to aid comparison.
        (b) 
        Upper bounds on achievable channel amplitude, indexed by channel number, for metasurface-like configurations with varying component size and separation distance. 
        The surface area and thickness of both the sender and receiver scale with the mediator for each case. 
        Unannotated lines for the (unstructured) vacuum comparisons correspond exactly with annotated separations for the bounds. 
        A common schematic showing the different dimensions of the sender, mediator and receiver for all four panels is shown in the top-left panel.}
    \end{center}
\end{figure*}
\noindent
Five applications of indexed channel amplitude bounds are explored as validation of the utility of the approach. 
In each case---upgrading prior results, two generalized studies exploring maximal three-dimensional transfer characteristic, a two-dimensional application mimicking the problem of laser awareness, and a final example investigating flux factors between finite three-dimensional objects---CFW bounds are either explicitly or implicitly verified to be reasonably tight: in cases where inverse designs are not provided the performance of unoptimized designs (vacuums and balls respectively) makes a strong case that the results are relatively tight.
Adding further constraints as dictated by need and circumstance, e.g. Ref.\;\cite{molesky2020hierarchical}, the evidence presented below suggests to us that CFW bounds can faithfully mirror near-optimal physics for many problems of current interest. 
\subsection{Applications: Rank Removal Upgrade}
\label{sec:rankUpgrade}
\noindent
The strengthening of the Frobenius norm bounds given by Eq.\;\eqref{eq:oldBoundsFrobenius} to equivalent per-channel results follows as a simple consequence of rank removal.  
Via Eq.\;\eqref{eq:WInvertedA} and the two forms of the CFW theorem
\begin{align}
    &\sigma_{n}\left(\bmm{W}_{rs}\right)\leq\,
    \sigma_{1}\left(\bmm{T}_{rr}^{\left(i\right)}\bmm{G}^{\circ}_{rs}\bmm{1}_{n-1}^{\perp}\bmm{W}_{ss}\right)=
    \nonumber \\
    &\left(\sigma_{1}\left(\bmm{W}_{ss}^{\dagger}\bmm{1}_{n-1}^{\perp}\left(\bmm{G}^{\circ\dagger}_{sr}\bmm{T}_{rr}^{\left(i\right)\dagger}\bmm{T}_{rr}^{\left(i\right)}\bmm{G}^{\circ}_{rs}\right)\bmm{1}_{n-1}^{\perp}\bmm{W}_{ss}\right)\right)^{1/2}.
    \nonumber
\end{align}
That is, if $\bmm{1}_{n-1}^{\perp}$ is selected to be a subspace perpendicular to the first $n\!-\!1$ right singular vectors of $\bmm{G}^{\circ}_{rs}$, then bounds on $\sigma_{n}\left(\bmm{P}_{rs}\right)$ follow via Eq.\;\eqref{eq:oldBoundsFrobenius} by considering a length one partial trace beginning $k = n$.
As such, the results of Eq.\;\eqref{eq:oldBoundsFrobenius} apply for each individual index $k$.
\\ \\
Incidentally, this same result can be obtained more directly by reformulating the argument of Ref.\;\cite{molesky2020fundamental} as an optimization. 
Let $\left\{\xi_{k},\bm{x}_{k}\right\}$ be the eigensystem of $\bmm{X}_{ss}$ ($\bmm{X}_{ss} \!\! = \!\! \sum_{k}\xi_{k}\,\bm{x}_{k}\otimes\bm{x}_{k}^{\dagger}$) and $\bm{t}_{k} = \bmm{T}_{ss}\bm{x}_{k}$. 
Because $\bmm{X}_{ss}$ depends only on the structuring of the receiver, as a consequence of rank removal, the $n^{th}$ singular value of $\sigma_{n}^{2}\left(\bmm{P}_{rs}\right)$ is bounded by 
\begin{align}
    &\max_{\left\{x_{k}\left(\bm{\chi}_{r}\right)\right\}}\bigg(
    \max_{\left\{\bmm{t}_{k}\left(\bm{\chi}_{s}\right)\right\}} 
    \frac{4}{\zeta_{s}}\sum_{k}x_{k}^{2}\,\bm{t}_{k}^{\dagger}\bmm{X}_{ss}^{\left(n\right)}\bm{t}_{k}
    \label{eq:pOpt} \\
    &~\text{s.t.}\,\left(\forall k\right)\,
    \Im\left(\bm{t}_{k}^{\dagger}\bm{x}_{k}\right)\geq\bm{t}_{k}^{\dagger}\left(\bm{\zeta}_{ss}^{-1} + \bmm{X}_{ss}\right)\bmm{t}_{k},
    \nonumber\\ 
    &~~\,\&\,\left(\forall k\right)\,\sigma_{k}\left(\bmm{X}_{ss}\right)\leq\zeta_{r}\,\sigma_{k}^{2}\left(\bmm{G}^{\circ}_{rs}\right)\bigg)\,\text{s.t.}\,\sum_{k}x_{k}^{2} = 1,
    \nonumber
\end{align}
where $\left\{x_{k}\right\}$ is the set of expansion coefficients of the CFW optimization vector in the $\bmm{X}_{ss}$ basis, $\bm{\zeta}_{ss}^{-1} = \zeta_{s}^{-1}\bmm{1}_{ss}$, $\bmm{X}^{\left(n\right)}_{ss} = \sum_{k = n}\xi_{k}\,\bm{x}_{k}\otimes\bm{x}_{k}^{\dagger}$, and the last set of inequalities following from the constraint of maximal material response\;\cite{miller2016fundamental}.
Without imposing any additional cross-constraints between the distinct $\bm{t}_{k}$ images---which in reality must be correlated since they are all situated in a single material structure\,\;\cite{molesky2021comm}---Lagrange duality shows that the inner optimizations decompose over $k$ as
\begin{align}
    &\max_{\left\{x_{k}\right\}}
    \sum_{k}x_{k}^{2}\,
    \begin{cases}
        \,~~~~~~~\,0 & k < n, \\
        \,~~~~~~~\,1 & k\geq n~\&~\kappa_{k} \geq 1 \\
        \,4\kappa_{n}/\left(1 + \kappa_{n}\right)^{2} & k\geq n~\&~\kappa_{k}\leq 1
    \end{cases},
    \nonumber \\
    &~\text{s.t.}\,\sum_{k}x_{k}^{2} = 1,
    \nonumber
\end{align}
where $\kappa_{n} = \zeta_{s}\zeta_{r}\,\sigma_{n}^{2}\left(\bmm{G}^{\circ}_{rs}\right)$. 
Maximizing over $\left\{x_{k}\right\}$,
\begin{equation}
    \sigma_{n}^{2}\left(\bmm{P}_{rs}\right)\leq
    \begin{cases}
       ~~~~~~~~~\,\,1 & \kappa_{n} \geq 1 \\
        4\kappa_{n}/\left(1 + \kappa_{n}\right)^{2} & \kappa_{n}\leq 1
    \end{cases},
    \label{eq:anaResultPrsAmp}
\end{equation}
with associated bounds on $\bmm{G}_{rs}$ and $\bmm{W}_{rs}$ following via Eq.\;\eqref{eq:channelAmplitudeConversions}\,\footnote{To use Eq.\;\eqref{eq:anaResultPrsAmp} in the presence of a mediator, one needs to decide which object the mediator will be grouped into, and then apply Eq.\;\eqref{eq:anaResultPrsAmp} with the best material parameters. 
Analogous results also hold for sources and receivers potentially composed of multiple materials: Eq.\;\eqref{eq:anaResultPrsAmp} remains applicable so long as the best $\chi_{m}$ factor is used for each volume.}.
Like the vacuum separation results, which appear as the limiting form when $\kappa_{n}\ll 1$, Eq.\;\eqref{eq:anaResultPrsAmp} indicates that the singular value decay of the vacuum separation is unavoidable. 
At the same time, Eq.\;\eqref{eq:anaResultPrsAmp} also proves that multiple scattering and rate matching effects limit achievable channel amplitudes far below the bound of Eq.\;\eqref{eq:1NinqResults} when either $\zeta_{s}$ or $\zeta_{r}$ are $\gg 1$.
Photonic power transfer, regardless of material structuring, must respect Landauer like limits\;\cite{landauer1989johnson}. 
Irrespective of the scattering processes that are utilized, it is not possible to transfer more power than what is provided to the initial current distributions.
\\ \\
While we advocate for the use of more exact and systematic optimization forms going forward, Eqs.\;\eqref{eq:pOpt} and \eqref{eq:anaResultPrsAmp} will almost certainly remain a useful point of reference. 
Using randomized singular value decomposition (Ref.\;\cite{halko2011finding} or Supporting Material) or analytic forms\;\cite{venkataram2019fundamental,miller2025tunnelling}, approximate vacuum channel amplitudes are surprisingly accessible for many photonic device design problems. 
Once this information is obtained, Eq.\;\eqref{eq:anaResultPrsAmp} is automatic. 
However, in employing Eq.\;\eqref{eq:anaResultPrsAmp} over heavier alternatives one should always begin with the belief that the results will be loose. 
First, as the size of the design reaches approximately wavelength scale, the findings presented below indicate that the value of $\kappa_{n}$ will likely somewhat over predict the number of channels that can reach saturation. 
Second, since Eq.\;\eqref{eq:pOpt} does not constrain relative phase response, prior literature all but guarantees that systems where it is difficult to form a large number of resonances (e.g. small domains with weakly dielectric media) will also be problematic\;\cite{molesky2020t}.
\subsection{Applications: Waveguides and Metasurfaces}
\noindent
As waveguides provide one of the archetypical examples of communication via channels---by extruding some material pattern, power can be efficiently coupled between a pair of (largely) separated volumes through a small set of discrete, low-loss modes---our computational applications begin by examining channel bounds in the context of waveguide like mediators.
\\ \\
Fig.\;\ref{fig:svdBounds} (a) bounds the number of independently addressable channels a photonic system of given size can support above a chosen cutoff threshold $c$: e.g., focusing on the upper left frame, no matter what structuring is employed, no $1.0\times 1.0\times 5.0\,\lambda$ mediator derived from a silicon-like material ($\chi = 13.6 + 0.05i$) can support more than twenty-five $\bmm{G}_{rs}$ channels with unit amplitude. 
One of the most interesting aspects of this data is the somewhat astounding extent to which the bounds allow for the possibility of spatially multiplexing medium distance communication using a wavelength scale structure.
Towards the $l = 5\lambda$ right edge of the upper plot, all but the $c = 0.10$ cutoff line appear to be tending towards near stable asymptotics. 
Although the appearance of these plateaus may in part be due to numerical artifacts, the recorded channel counts are significantly larger than what is realized in typical waveguides. 
\\ \\
The metasurface-style limits depicted in Fig.\;\ref{fig:svdBounds} (b) apply to the complementary situation of transmitting information across some fixed distance using a comparatively thin mediator: e.g., beam shaping\;\cite{yang2018freeform}, mode conversion\;\cite{akgol2018design}, and (de)multiplexing\;\cite{frellsen2016topology}.
Several noteworthy features can be identified across the four panels. 
First, as a function of the side length parameter $h$, the number of $\bmm{G}_{rs}$ channels above a given cutoff is seen to rapidly increase and then effectively saturate; in going from $1.0\times 1.0\times 1.0\,\left(\lambda\right)$ to $4.0\times 4.0\times 1.0\,\left(\lambda\right)$ the maximal number of realizable channels having at least unity amplitude at a separation distance of $d = \lambda/ 4$ grows from $\approx 8$ to $\approx 32$, exhibiting scaling roughly proportional to the area; from  $4.0\times 4.0\times 1.0\,\left(\lambda\right)$ to $8.0\times 8.0\times 1.0\,\left(\lambda\right)$ only minor changes in the bounds occur until the separation reaches $\approx 5\lambda$, which may be simply attributed to the larger solid angle covered by the larger components at a fixed distance. 
Second, as indicated by the relative differences between the $4.0\times 4.0\times 4.0\,\left(\lambda\right)$ bounds and the $8.0\times 8.0\times 1.0\,\left(\lambda\right)$ bounds, the ratio of thickness to area plays an important role in setting  channel amplitude limits. 
For equal design volumes, the bounds intimate that thicker components allow for the realization of more large amplitude channels until a separation distance of $\approx 5.0\lambda$. 
Noting that the thickness of the mediator is linked to the thickness of the sender and receiver regions, it is not possible to disentangle whether these difference are primarily an effect of the distinct device geometries offered by altering the mediator design volume, or whether they are mainly caused by changes in the types of fields that can be excited and detected based on Fig.\,2(b). 
Nevertheless, regardless of the precise origin, increasing thickness near the wavelength scale produces larger channel amplitude limits\;\cite{miller2023optics}. 
\\ \\
While care should always be taken in inferring absolute design lessons from limits---the bounds may simply be loose---the suggested implications for metasurfaces are both sensible and non-trivial. 
At any given separation, there is likely much to be gained in regards to achievable multi-functionality by designing over larger ``non-local'' cell volumes: at first, increasing the design area greatly increases the number of useable channels. 
Yet, these improvements do not persist indefinitely, at least as far as can be extrapolated from channel bounds on the relatively small mediators we have studied, and at a certain point thickness becomes the limiting factor. 
\subsection{Applications: Laser Awareness (Fisher Information)}
\label{sec:Fisher}
\begin{figure}[htp!]
    \includegraphics[width=\linewidth]{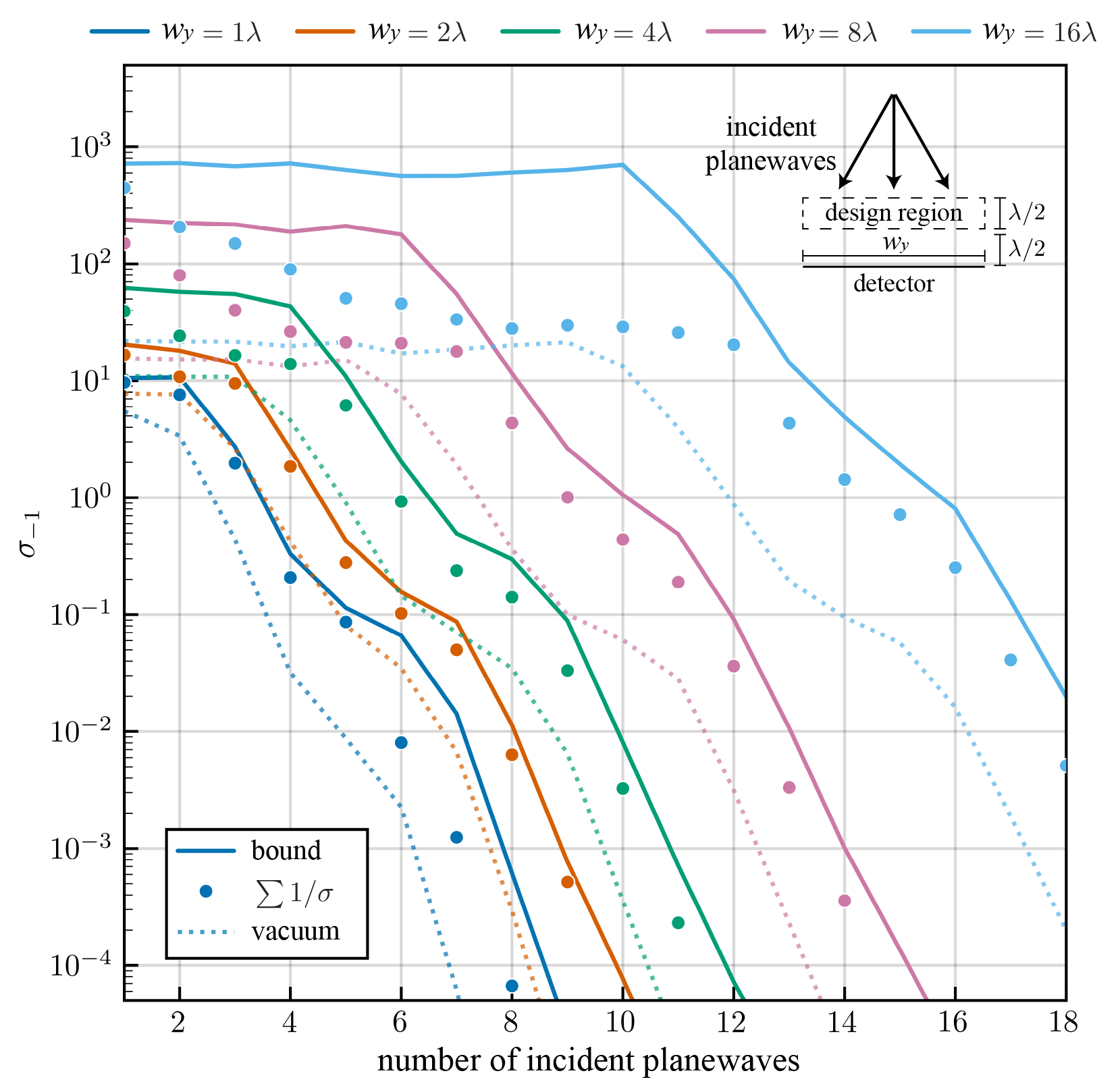}
    \caption{\label{fig:laserAwareness} 
    \textbf{Planewave angle discrimination.}
    The inset depicts a schematic of the planewave angle detection metasurface design scenario considered in the plot. 
    The metasurface design region has dimensions $0.3\lambda\times w_{y}$. 
    The detection region is a line segment of length $w_{y}$, separated from the metasurface by $\lambda/2$, and the input space is spanned by a basis of $n$ planewaves with wavevectors at angles $\phi_l = \phi_{min} + (l-1) (\phi_{max}-\phi_{min})/(n-1) $. 
    The main plots shows bounds on $\sigma_{-1}(\bmm{C})$, and inverse designs maximizing $\sigma_{-1}(\bmm{C})$, for incident planewave detection, with $\phi_{min/max} = \pm 15^{\circ}$, for various detector widths and number of incident planewaves.
    }
\end{figure}

\noindent
Bounds on the minimum singular value $\sigma_{-1}$, over some set of a fixed size, are often useful in the context of photonic information processing systems. 
As a more application oriented study, we now consider a set of channels as a stochastic linear transformation $\bm{y} = \bmm{C}\,\bm{x} + \bm{n}$, with input $\bm{x}$, output $\bm{y}$, linear channel matrix $\bmm{C}$, and additive noise $\bm{n}$. 
Practical examples of such systems include waveguide multiplexers where $\bm{x}$ and $\bm{y}$ are the modal coefficients of the input and output waveguides, as well as on chip spectrometers\;\cite{ma2025inverse}. 
\\ \\
One of the most common goals encountered in information based applications is to engineer $\bmm{C}$ (usually the Green operator or $W$-operator) in order to induce robustness to noise as qualified by the condition number $\text{cond}\left(\bmm{C}\right) = \sigma_{1}\left(\bmm{C}\right)/\sigma_{-1}\left(\bmm{C}\right)$. 
Intuitively, $\text{cond}\left(\bmm{C}\right)$ is a measure of the sensitivity of the inferred input $\bm{x}$ to the observed noisy output $\bm{y}$. 
Minimizing $\text{cond}\left(\bmm{C}\right)$, trivially, amounts to minimizing $\sigma_{1}\left(\bmm{C}\right)$ and maximizing $\sigma_{-1}\left(\bmm{C}\right)$. 
However, minimizing $\sigma_{1}\left(\bmm{C}\right)$ is often undesirable, as the largest channel amplitude directly corresponds to the maximum strength of a transmitted signal, which should (naively) be kept as high as possible in most circumstances.
Consequently, $\sigma_{-1}\left(\bmm{C}\right)$ is frequently the critical driver of $\text{cond}\left(\bmm{C}\right)$, and upper bounds on $\sigma_{-1}\left(\bmm{C}\right)$ effectively capture the scaling of noise robustness with regards to design parameters such as material susceptibility and device footprint for a given number of channels.
\\ \\
Another perspective on the role of $\sigma_{-1}\left(\bmm{C}\right)$ is provided by the Fisher information. 
Supposing additive Gaussian noise $\bm{n}\sim\mathcal{N}\left(0,n\right)$, so that $\bm{y}\sim\mathcal{N}\left(\bmm{C}\bm{x},n\right)$, the Fisher information matrix of $\bm{y}$ about $\bm{x}$ is $I_{\bm{y}}\left(\bm{x}\right) = \bmm{C}^\dagger\bmm{C}/n$. 
Taking measurement to be formally described as an estimation of the value of $\bm{x}$ given the observed $\bm{y}$, $I_{\bm{y}}\left(\bm{x}\right)$ sets the minimum estimator covariance via the Cram\'er-Rao bound: for any unbiased estimator $\text{est}\left(\bm{y}\right)$ of $\bm{x}$, $\text{cov}\left(\text{est}\left(\bm{y}\right)\right)\succeq I_{\bm{y}}\left(\bm{x}\right)^{-1} = n\,(\bmm{C}^\dagger\bmm{C})^{-1}$. 
Because less estimator covariance means more estimation certainty, it is desirable to make $(\bmm{C}^\dagger\bmm{C})^{-1}$ small in some sense, and maximizing $\sigma_{-1}\left(\bmm{C}\right)$ is equivalent to minimizing the spectral norm of $(\bmm{C}^\dagger\bmm{C})^{-1}$. 
$\sigma_1(\bmm{C})$, for this metric, has no significance. 
\\ \\
As a demonstration of CFW bounds for $\sigma_{-1}(\bmm{C})$, Fig.\;\ref{fig:laserAwareness} examines incident planewave discrimination via a structurable metasurface mediator and designable detector (receiver), depicted schematically in Fig.\;\ref{fig:laserAwareness} (top). 
In greater detail, electric fields within the detector region are used to infer a decompositions of the incident fields in terms of planewaves along different directions: $\bmm{C} = \bmm{G}_{rf}\bmm{S}_{ff}$, where the columns of $\bmm{S}_{ff}$ are unit planewave sources in the far-field with distinct incident angles $\phi$. 
The metasurfaces $\epsilon\left(\bm{r}\right) = \left\{6 + 0.01i,\, 1\right\}$ are designed via topology optimization to discover devices with large discrimination capabilities. 
Likely due to the fact that $\sigma_{-1}(\bmm{C})$ is not differentiable when $\bmm{C}$ has degenerate minimal singular vectors, directly maximizing $\sigma_{-1}(\bmm{C})$ was found to be challenging, and the proxy objective of minimizing $\sum 1/\sigma_j$ (Ref.\;\cite{ma2025inverse}) was substantially more effective. 
Comparative bounds on $\sigma_{-1}(\bmm{C})$ are computed via subspace optimization: for each probing $\bm{s}$, quadratically constrained quadratic program (QCQP) upper bounds are computed on $\bm{s}^\dagger\bmm{C}^\dagger\bmm{C}\bm{s}$, which always bounds $\sigma_{-1}^2(\bmm{C})$, and the tightest such bound is then found by minimizing over $\bm{s}$ (see Supplementary Material). 
\\ 
Bounds on $\sigma_{-1}(\bmm{C})$ exhibit a plateau with regards to the number of input planewaves less than some transition value, after which they decay exponentially. 
For a given detector width $w_{y}$, this indicates a threshold dimensionality of the space of input planewave combinations beyond which the capability of any metasurface to distinguish between different inputs degrades. 
The threshold is seen to increase roughly linearly with $w_{y}$, reflecting an increase in the number of usable channels between the far-field and the detector as mediated by an optimal metasurface. 
\subsection{Applications: Heat Transfer (Flux Factor)}
\label{sec:heatTransfer}
\begin{figure*}[htp!]
    \includegraphics[width=\linewidth]{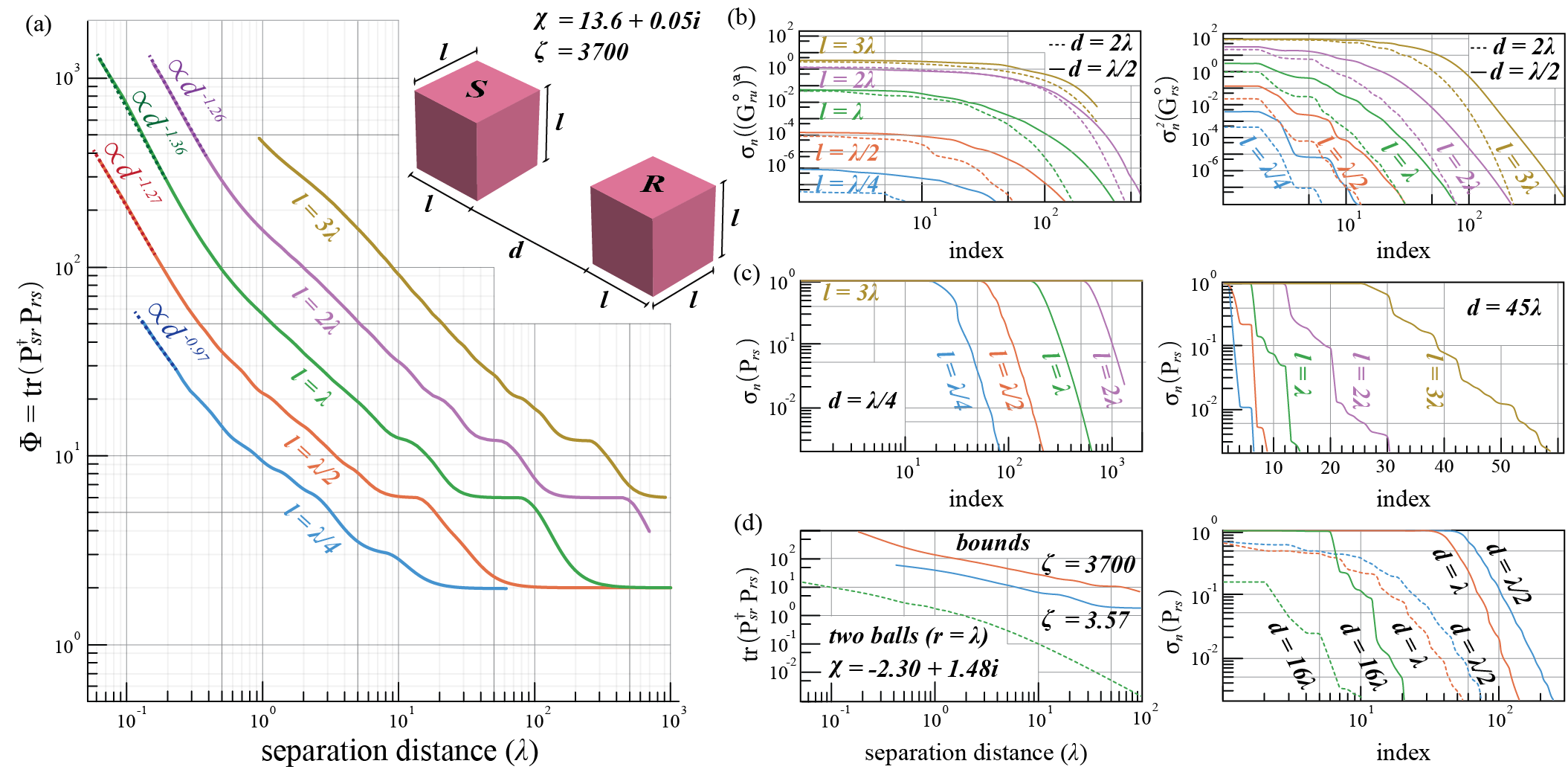}
    \caption{\label{fig:traceBounds} 
    \textbf{Limit flux factors for finite three-dimensional objects.}
    The figure depicts CFW bounds on the flux factor $\Phi_{rs} = \tr\left(\bmm{P}_{sr}^{\dagger}\bmm{P}_{rs}\right)$ appearing in radiative heat transfer, and associated indexed channel amplitude bounds, for the configuration inset of (a): cubic sender and receiver volumes of characteristic size $l$ (bounding domains for arbitrarily structured source and receiver geometries), separated by a distance $d$. 
    (a) Limit flux factors as function of $d$ and $l$ with approximate near-field asymptotics (where available), for a silicon-like material with $\chi = 13.6 + 0.05\,i$ ($\zeta = \left|\chi\right|^{2}/\Im\left(\chi\right) = 3700$). 
    (b) 
    Vacuum channel amplitudes used in computing (a) as function $l$ and index for two representative separations, see Eq.\;\eqref{eq:anaResultPrsAmp} and Supplementary Materials. 
    (c) 
    Indexed channel amplitude bounds for the flux factors given in (a) at $d = \lambda / 4$ and $d = 45\lambda$.
    (d) 
    The left panel provides a comparison between flux factor bounds and the value of $\Phi_{rs}$ achieved between wavelength radius silica balls supporting surface phonon-polariton resonances ($\chi = -2.30 + 1.48\,i$ at $20\,\mu\text{m} = 0.06\,\text{eV}$), solid lines (bounds) versus dashed line (balls). 
    The lower blue $l = \lambda$ line is for $\chi= -2.30 + 1.48\,i$ ($\zeta = 3.57$) matching the considered value of silica.
    The upper orange $l = \lambda$ line is for $\chi = 13.6 + 0.05\,i$ ($\zeta = 3700$) as depicted in (a). 
    The right panel decomposes the flux factor results from the left panel at three separation distances following the same line identification scheme as the left panel. 
    }
\end{figure*}
\noindent
As our final application of the CFW method, we return to the question of upper limits on radiative heat transfer previously examined in Refs.\;\cite{venkataram2020fundamental,molesky2020fundamental}.
Because of the unit amplitude bound on $\bmm{P}_{rs}$ given by Eq.\;\eqref{eq:anaResultPrsAmp}, and the rapid amplitude decay characteristics observed universally across various size and separation distance combinations (Fig.\;\ref{fig:traceBounds} (b) and (c)), computed bounds on the flux factor $\Phi$ effectively doubles as count bounds on the number of channels in saturation. 
\\ \\
Fig.\;\ref{fig:traceBounds} (a) illustrates $\Phi_{rs} = \tr\left(\bmm{P}_{sr}^{\dagger}\bmm{P}_{rs}\right)$ flux factor bounds---giving bounds on heat transfer via $h\left(\omega\right) = \Theta\left(\omega, T_{r}, T_{s}\right)\Phi_{rs}\left(\omega\right)$, \emph{Formulation: Power Scattering}---between finite three-dimensional sender and receiver objects restricted to cubic bounding domains. 
Counterintuitively, even at ``mid-field'' distances ($d\approx 10\lambda$) the bounds do not exhibit the area scaling trend that might be expected based on the asymptotic of geometric optics. 
Below $l = 2\lambda$, the scaling factor between different $l$ lines slowly accelerates up to a factor of $\approx 5/2$, which is well below the factor of $4$ area ratio. 
Between $l = 2\lambda$ and $l = 3\lambda$, a similar gap, fluctuating slightly with separation, is observed despite the fact that the area ratio is now only $9/4$. 
\\ \\
Transitioning to the near-field ($d\ll\lambda)$, $\Phi_{rs}$ bounds on larger domains scale as approximately $\propto d^{-5/4}$, which is faster than the $\propto d^{-1}$ separation scaling of balls\;\cite{narayanaswamy2008thermal}, but slower than the $\propto d^{-2}$ separation dependence of half-spaces and thin-films\;\cite{polder1971theory}. 
As an immediate consequence, in contrast to the conclusions of Ref.\;\cite{venkataram2020casimir} (see also Ref.\;\cite{zhang2020optimal}) that film geometries supporting polariton resonances are effectively optimal in the case of unbounded domains separated by a fixed vacuum gap, it follows that neither of these simple geometries saturates the CFW bound\;\footnote{That is, considering only the restriction of global real power conservation, see Ref.\;\cite{chao2022physical} and Supplementary Materials for additional details.}---the direct comparison between wavelength radius silica balls and the computed $\Phi_{rs}$ bounds given Fig.\;\ref{fig:traceBounds} (d) (lower solid blue curve $\zeta = 3.57$ bound curve versus dashed green curve of silica balls) shows a persistent gap in excess of one order of magnitude across all tested separations, caused by both sub-bound per-channel saturation and a lower index onset of exponential channel amplitude decay. 
\\ \\ 
Moving to larger separation distances, the trend first seen in Fig.\;\ref{fig:svdBounds} of CFW bounds not precluding spatial multiplexing well in excess of known device geometries is continued. 
For example, the $l = 3\,\lambda$ domain results for $\chi = 13.6 + 0.05\,i$ ($\zeta = 3700$) allow support for up to $\approx 20$ unit amplitude channels up a separation distance of $d = 100\,\lambda$, and continue to allow support of up to $6$ near-unit amplitude channels (likely related to dipole modes) to the maximum considered separation of $d = 1000\,\lambda$. 
Connecting back to questions concerning communication and antenna design, whether or not the plateau characteristics responsible for these larger bound values persist under the imposition of additional constraints, as they did in the laser awareness example, and if so whether they are achievable with some relatively simple design, rests as an important direction for further study. 
\section{\S Outlook}
\noindent
The most intriguing aspect of the CFW method is how well it appears to predict realizable functionality. 
For the three-dimensional examples, calculated results are similar enough to vacuum or basic ball geometries that, barring cases where the channel amplitudes are too small to be of practical interest, duality gaps substantially larger than an order of magnitude are unlikely (if not impossible) for optimized designs. 
For the laser awareness example, topology optimization generally finds designs within a factor of unity of the computed bounds, and in the situations in which larger differences are seen there are plausible explanations as to why such behaviour could be expected.
\\ \\ 
While the selection of a $Q^{n-1}$ subspace in the CFW formulation implies that distinct channels can not fully reuse a single resonance---inputs and outputs must remain orthogonal---consistency of the implied material profile is not explicitly enforced in any of the results we have presented. 
That is, although methods for asserting that each response is compatible with a single structure within the QCQP framework are available, e.g.\;\cite{molesky2021comm}, one of the primary benefits of the current implementation is the ability to jump to bounds on the $n^{th}$ channel amplitude without having to impose any relations with the previous $n$ channels\,\footnote{This independence also implies that various indicies can be computed in parallel without any communication overhead, which is critical to many of our three-dimensional results.}. 
If structural consistency constraints were included, one could imagine that the bounds given in \emph{Applications} might be reduced by significant factors, particularly when multiple large amplitude channels are accessible (the idea being that simultaneously activating many large amplitude channels in a single device of a moderate size will inescapably lead to performance reducing tradeoffs). 
Relatedly, in the three-dimensional examples there is no optimization over the exclusion subspace $Q^{n-1}_{\perp}$, Eq.\;\eqref{eq:CFWminmax}. 
The depicted bounds, as detailed further in Supporting Material, are the result of an educated guess for $Q^{n-1}$, attempting to single out the extent to which a scattering medium can influence vacuum decay characteristics. 
For bounds on $\bmm{G}_{rs}$, this guess almost certainly leads to some persistent overestimation of the attainable channel amplitude at large channel indices: to achieve the tightest bounds, $Q^{n-1}$ should be selected to balance $\bmm{G}^{\circ}$ against $\bmm{G}^{\circ}\bmm{T}\bmm{G}^{\circ}$ in Eq.\;\eqref{eq:gtExp} and this has not been done\,\footnote{Similar to the inclusion of structural consistency constraints, this shortcoming could also be mitigated at the cost of additional computational effort.}.
\\ \\
The choice to ignore these realities of photonic channels makes the agreement observed in \emph{Applications} all the more interesting. 
Setting aside the physical context, there is no persuasive reason to suppose that use of the CFW theorem would lead to practically tight indexed singular value bounds in the way that we have employed it. 
In our CFW formulation, the (outer) excluded subspace $Q^{n-1}_{\perp}$ is always selected prior to the (inner) structural optimization from which bounds are determined. 
Supposing the existence of at least two devices (operators) capable of mapping multiple, distinct, orthogonal inputs to multiple, distinct, orthogonal outputs, it is easy to imagine plausible failure modes. 
For example, consider a set of devices $\left\{\bmm{W}_{1},\ldots,\bmm{W}_{n}\right\}$ such that, for all $1\leq k\leq n$, $\sigma_{1}\left(\bmm{W}_{k}\right)\approx v$.
If the first channel of each $\bmm{W}_{k}$ is orthogonal to the first channel of any other $\bmm{W}_{j}$, then the CFW approach could at best report a nearly constant bound of $\approx v$ for all $1\leq k\leq n$, even if in reality each device can only support one channel amplitude near $v$.
\\ \\ 
The fact that decay characteristics seemingly consistent with a single device are nevertheless observed---Eq.~\eqref{eq:anaResultPrsAmp} and Ref.\;\cite{kuang2025bounds} analytically prove that exponential decay always occurs beyond some index, which is incongruous with the hypothetical multi-device failure mode presented above---reinforces the notion that the physics of scattering plays a central role in the viability of Eq.\;\eqref{eq:CFWminmax}. 
Because of the inseparable tether to vacuum physics, arbitrary material structuring can not create arbitrary channel characteristics\;\cite{kuang2025bounds}. 
Echoing the conclusions of other recent works on fundamental limits in device engineering\;\cite{chao2022physical,guo2024passivity}, we take this as convincing evidence that the mathematics of scattering is both deeper and more restrictive than it superficially appears. 
\\ \\
Looking forward, there are a number of substantial extensions and applications of the CFW approach that we have not explored in this work. 
How do channel amplitude bounds change when applied to finite frequency windows\;\cite{liang2013formulation,shim2019fundamental,zhang2023all}? 
Does the finding that seemingly reliable bounds can be computed without cross-constraints have implications for multi-functional device design\;\cite{gertler2025many,chao2025blueprints}? 
Are the suggestions of greatly enhanced spatial multiplexing seen in \emph{Applications} actually realizable? 
Beyond these questions, we also highlight that indexed channel amplitude bounds (for large values of $n$) may be quite useful for establishing convergence guarantees, error limits, and pre-allocating memory in computational methods involving decomposition and reduced rank approximations\;\cite{chew1993nepal,martinsson2007fast,liu2021butterfly,xue2023fullwave,mavrikakis2025surface}.
\\ \\
As a closing comment on utility, note that in comparison with existing QCQP convex relaxation bounds, see Refs.\;\cite{chao2022physical,miller2026fundamental}, CFW channel amplitude bounds are substantially more general and flexible. 
In essence, whenever a quadratic (or sub-quadratic) objective can be stated in terms of linear scattering operators the CFW based approach we have presented here is applicable, without any need need to find an explicit duality transformation form. 
Further, because of the special self-similar algebraic structure that appears in CFW channel amplitude bounds, in contrast to one of the central lessons of inverse design, there is a distinct chance that (in certain instances) indexed singular value bounds may lead to tighter results than working with the actual objective of interest\;\cite{chakravarthi2020inverse,christiansen2023impact}. 
\section{Acknowledgements}
\noindent
A.W. Rodriguez and A. Amaolo acknowledge support by a Princeton SEAS Innovation Grant and by the Cornell Center for Materials Research (MRSEC). 
S. Molesky and P. Virally acknowledge financial support from NSERC under Discovery Grant RGPIN-2023-05818 and the CGS-M program, as well as additional benefits provided from their affiliations to the Regroupement Québécois sur les Matériaux de Pointes, https://doi.org/10.69777/309032, the IVADO Research Consortium, and the Lassonde Deeptech Institute. 
The simulations presented in this article were performed on computational resources managed and supported by Calcul Québec (calculquebec.ca) and the Digital Research Alliance of Canada (alliancecan.ca) under resource allocation RRG \#5398, as well as the MIT SuperCloud and Lincoln Laboratory Supercomputing Center.
\\ \\
\begin{center}
    \textbf{\large Supporting Material}
\end{center}
The supporting material is divided into three sections: \emph{Formulation Support}, \emph{Computational Expedients}, and \emph{Numerical Methods}. 
The first section reproduces the von Neumann ``basis chaining'' inequality used in Ref.\;\cite{molesky2020fundamental} to derive Frobenius norm bounds on $\bmm{P}_{rs}$ and provides additional details on the CFW optimization forms relevant to determining indexed singular value bounds for electromagnetic scattering operators. 
Building off this discussion, \emph{Computational Expedients} then describes how CFW programs can be simplified and approximated to mitigate computational costs. 
At present, these reductions are essential for treating wavelength scale three-dimensional design problems on workstation level systems. 
The final section provides an overview of the numerical methods used to produce the data presented in Figs.\;\ref{fig:svdBounds}--\ref{fig:laserAwareness}.
\section{\S Formulation Support}
\label{sec:Formulation_Support}
\subsection{Von Neumann Trace Inequality}
\noindent
We begin by providing a proof of the von Neumann trace inequality. 
As indicated in Ref.\;\cite{molesky2020fundamental}, the trace inequality (coupled with rank removal) provides the basis of an alternate means of reaching the analytic singular value bounds on $\bmm{P}_{rs}$ given by Eq.~\;\eqref{eq:anaResultPrsAmp}. 
\\ \\
\emph{Trace Inequality}: Let $\bmm{A}:X\rightarrow Y$ and $\bmm{B}:Y\rightarrow X$ be linear operators between the complex vector spaces $X$ of dimension $l$ and $Y$ of dimension $m$. 
Take $\bmm{A} = \bmm{V}_{a}\bmm{D}_{a}\bmm{U}_{a}^{\dagger}$ and $\bmm{B} = \bmm{V}_{b}\bmm{D}_{b}\bmm{U}_{b}^{\dagger}$ to be SVDs of $\bmm{A}$ and $\bmm{B}$. 
$$
    \left|\tr\bmm{B}\bmm{A}\right|\leq\sum^{\min\set{l,m}}_{k=1}\sigma^{b}_{k}\sigma^{a}_{k},
    \label{eq:trcInq}
$$
where $\left\{\sigma_{1}^{a},\ldots,\sigma_{l}^{a}\right\}$ is the set of singular values of $\bmm{A}$ and $\left\{\sigma_{1}^{b},\ldots,\sigma_{m}^{b}\right\}$ is the set of singular values of $\bmm{B}$. 
\begin{proof}[Proof of Trace Inequality]
    Assume that $\bmm{B}$ is rank one, $\bmm{B} = \sigma^{b}\bmm{v}_{b}\bmm{u}_{b}^{\dagger}$ with $\bmm{v}_{b}$ and $\bmm{u}_{b}$ unit vectors in $X$ and $Y$ and $\sigma^{b} > 0$. 
    Take $c_{1},...,c_{l}$ to be the coefficients of expansion of $\bmm{v}_{b}$ in the basis of $\bmm{U}_{a}$. and $d_{1},...,d_{m}$ to be the coefficients of expansion of $\bmm{u}_{b}$ in the basis of $\bmm{V}_{a}$. 
    \begin{align}
        &\left|\tr\bmm{B}\bmm{A}\right| 
        = \sigma^{b}\left|\sum_{k,j,i}\sigma^{a}_{k}\left(\bmm{v}^{a\dagger}_{i}\bmm{v}^{a}_{k}\right)\left(\bmm{u}_{k}^{a\dagger}\bmm{u}_{j}^{a}\right)\left(c_{j}d_{k}^{*}\right)\right| =
        \nonumber \\
        &\sigma^{b}\left|\sum_{k}\sigma^{a}_{k}c_{k}d_{k}^{*}\right|
        \leq
        \sigma^{b}\sqrt{\left|\sum_{k}\sigma^{a}_{k}c_{k}c_{k}^{*}\right|\,\left|\sum_{k}\sigma^{a}_{k}d_{k}d_{k}^{*}\right|}\leq
        \nonumber \\
        &\sigma^{b}\sigma_{1}^{a}\sqrt{\left|\sum_{k}c_{k}c_{k}^{*}\right|\,\left|\sum_{k}d_{k}d_{k}^{*}\right|} = \sigma^{b}\sigma^{a}_{1},
        \nonumber
    \end{align}
    with the third relation following from the Cauchy-Schwarz inequality. 
    Setting $c_{1} = d_{1} = 1$, $\left|\tr\bmm{A}\bmm{B}\right| = \sigma^{a}_{1}\sigma^{b}$. 
    \\ \\
    Suppose that the proposition has been verified for the composition of any rank $n-1$ linear operator with $\bmm{A}$. 
    Let $\bmm{B} = \bmm{E} + \bmm{F}$, with $\bmm{F} = \sigma^{b}_{n}\bmm{V}_{b}\bmm{U}_{b}^{\dagger}$ and $\bmm{E} = \bmm{V}_{b}\left(\bmm{D}_{b}-\bmm{1}_{n}\sigma_{n}^{b}\right)\bmm{U}_{b}^{\dagger}$. 
    Here, $\bmm{1}_{n}$ is the identity operator with zeros replacing the usual ones at all positions along the main diagonal greater than $n$---the identity restricted to the first $n$ singular values. 
    Extending the conventions used above, 
    \begin{align}
        \left|\tr\bmm{F}\bmm{A}\right| 
        &= \sigma^{b}_{n}\left|\sum_{q}\sum_{k,j,i}\sigma^{a}_{k}\left(\bmm{v}^{a\dagger}_{i}\bmm{v}^{a}_{k}\right)\left(\bmm{u}_{k}^{a\dagger}\bmm{u}_{j}^{a}\right)\left(c_{j}^{q}d_{k}^{q*}\right)\right|
        \nonumber \\
        &= \sigma^{b}_{n}\left|\sum_{k}\sigma_{k}^{a}\left(\sum_{q}c_{k}^{q}d_{k}^{q*}\right)\right|\leq\sum_{k}\sigma^{b}_{n}\sigma^{a}_{k}\left|\sum_{q}c_{k}^{q}d_{k}^{q*}\right|
        \nonumber \\
        &\leq \sum_{k}\sigma^{b}_{n}\sigma^{a}_{k},
        \nonumber
    \end{align}
    with the third relation following from the triangle inequality, and the fourth relation following from the Cauchy-Schwarz inequality. 
    Therefore, using the fact that $\left|\tr\bmm{E}\bmm{A}\right|\leq\sum_{k}\sigma^{d}_{k}\sigma^{a}_{k}$ because $\bmm{E}$ is of rank $n-1$, 
    \begin{align}
        \left|\tr\bmm{B}\bmm{A}\right| &\leq\sum_{k}\sigma^{d}_{k}\sigma^{a}_{k} + \sum_{k}\sigma^{b}_{n}\sigma_{k}^{a} 
        \nonumber \\
        &= \sum_{k}\left(\sigma_{k}^{d}+\sigma_{n}^{b}\right)\sigma_{k}^{a} = \sum_{k}\sigma^{b}_{k}\sigma^{a}_{k}.
        \nonumber
    \end{align}
    Setting $\bmm{U}_{b} = \bmm{V}_{a}$ and $\bmm{V}_{b} = \bmm{U}_{a}$ the inequality is saturated, proving that these choices are optimal.  
\end{proof}
\noindent
As a consequence of the trace inequality argument, selecting $\bmm{U}_{b} = \bmm{V}_{a}$ and $\bmm{V}_{b} = \bmm{U}_{a}$ also acts to maximize every partial trace of $\left|\tr\bmm{B}\bmm{A}\right|$: 
$$
    \lVert\bmm{B}\bmm{A}\rVert_{\mathtt{F}_{n}}^{2}\leq\sum_{k=1}^{n}\sigma^{a}_{k}\sigma^{b}_{k},
$$
with $\lVert\bmm{B}\bmm{A}\rVert_{\mathtt{F}_{n}}^{2}$ denoting the Frobenius norm of $\bmm{B}\bmm{A}$ restricted to the subspace spanned by the first $n$ (left or right) singular vectors---the $n^{th}$ partial Frobenius norm.
\subsection{Optimization Forms: Green Operator}
\noindent
The form of the Green operator used to compute the mediator design results given in the main text is  
\begin{align}
    \bmm{G}_{rs} &= \bmm{G}^{\circ}_{rs} + \bmm{G}^{\circ}_{rd}\bmm{T}_{dd}\left(\bmm{\chi}\right)\bmm{G}^{\circ}_{ds},
    \label{eq:greenScattering}
\end{align}
where, as before, the ${vac}$ superscripts denotes a bare (vacuum) operator, $\bmm{T}_{dd}(\bmm{\chi})$ is the scattering $T$-operator---mapping incident electromagnetic fields to the associated total, self-consistent, polarization currents induced by the photonic system. 
The $d$ subscript indicates the designable part of the device, which is presently assumed to contain the mediator, and possibly contain some part of the receiver and source, c.f. Ref.\;\cite{amaolo2024maximum}.
For simplicity, $\bmm{\chi}$ is assumed to be a dielectric design profile $\epsilon\left(\bm{r}\right) - 1$: i.e. $\bmm{\chi}$ is effectively the variable of the optimization program
\begin{align}
    &\underset{\epsilon(\bm{r})}{\max/\min}\quad\sigma_{n}\{\bmm{G}_{rs}\}
    \nonumber\\
    &\,\text{s.t.}\,\left[k_{\circ}^{-2}\,\nabla\times\nabla\times\,\, - \epsilon\left(\bm{r}\right)\right]\bmm{G}\left(\bm{r},\bm{r'}\right) = \bmm{1}\delta(\bm{r}-\bmm{r'}) 
    \nonumber\\
    &\,\epsilon(\bm{r}) = 
    \begin{cases} 
        \text{fixed} & \bm{r} \notin D \\ 
        1 \text{ or } \chi & \bm{r} \in D 
    \end{cases}.
    \label{eq:structOpt}
\end{align}
(More generally, $\bm{\chi}$ can be used to describe to essentially any linear susceptibility\;\cite{kuang2020maximal}.) 
Although the dependence of $\bmm{T}$ on $\bmm{\chi}$ is present throughout, as it also is in the main text, it is hereafter suppressed to condense notation. 
Per Eq.\;\eqref{eq:CFWminmax}, let $\bmm{Q}_{\perp}$ be a column grouping of orthonormal basis vectors for $Q^{n-1}_{\perp}$ and $\bmm{Q}$ be a column grouping of orthonormal basis vectors for $Q^{n-1}$---$\bmm{Q}_{\perp} = \bmm{1} - \bmm{Q}$. 
Setting $\lVert\bm{s}\rVert_{2} = 1$, denote the polarization field induced by $\bm{q}_{\perp}\equiv\bmm{Q}_{\perp}\bm{s}$ within the structure as $\bm{t}=\bmm{T}_{dd}\bmm{G}_{ds}^{\circ}\bm{q}_{\perp}$. 
An upper bound on $\sigma_{n}$ for a fixed design then is given by
\begin{align}
    &\max_{\bm{q}_{\perp}}~\bm{q}_{\perp}^{\dagger}\bmm{G}_{rs}^{\circ\dagger}\bmm{G}_{rs}^{\circ}\bm{q}_{\perp} + 2\Re\left\{\bm{q}_{\perp}^{\dagger}\bmm{G}_{rs}^{\circ\dagger}\bmm{G}_{rd}^{\circ}\bm{t}\right\} + 
    \nonumber \\
    &~~~~~~~~\bm{t}^{\dagger}\bmm{G}_{rd}^{\circ\dagger}\bmm{G}_{rd}^{\circ}\bm{t}
    \nonumber \\
    &\,\text{s.t.}\,\lVert\bmm{Q}^{\dagger}\bm{q}_{\perp}\rVert_{2}^{2} = 0
    \text{\,\&\,}\lVert\bm{q}_{\perp}\rVert_{2}^{2} = 1.
    \nonumber
\end{align}
($\bm{t}$ is completely determined by $\bm{q}_{\perp}$ for a fixed structure.)
To obtain bounds on $\sigma_{n}$ applicable to all structures, the strict relation between $\bm{t}$ and $\bm{q}_{\perp}$ is relaxed to some subset of the structure-agnostic constraints set
\begin{equation}
    \left(\forall\bmm{R}^{i}\right)\,\bm{q}^{\dagger}_{\perp}\bmm{G}_{ds}^{\circ\dagger}\bmm{R}^{i}\bm{t} = \bm{t}^{\dagger}\left(\bm{\chi}^{-1} - \bmm{G}_{dd}^{\circ}\right)^{\dagger}\bmm{R}^{i}\bm{t}, 
    \label{eq:quadCstrts}
\end{equation}
where $\bmm{R}^{i}$ is a spatial ``resolution'' onto some subregion $R_{i}$ of the design $R$:
Eq.\;\eqref{eq:quadCstrts} asserts that real and reactive power must be conserved by any $(\bm{q}_{\perp},\bm{t})$ pair over $R_{i}$; a more detailed discussion of Eq.\;\eqref{eq:quadCstrts} can be found in Ref.\;\cite{chao2022physical}. 
Applying this relaxation, upper bounds on $\sigma_{n}$ for any design are given by
\begin{align}
    &\max_{\bm{q}_{\perp},\bm{t}}\,\bmm{q}^{\dagger}_{\perp}\bmm{G}_{rs}^{\circ\dagger}\bmm{G}_{rs}^{\circ}\bm{q}_{\perp} + 2\Re\left\{\bm{q}^{\dagger}_{\perp}\bmm{G}_{rs}^{\circ\dagger} \bmm{G}_{rd}^{\circ}\bm{t}\right\} + 
    \nonumber \\
    &~~~~~~~~\bm{t}^{\dagger}\bmm{G}_{rd}^{\circ\dagger}\bmm{G}_{rd}^{\circ}\bm{t} 
    \nonumber \\
    &\text{s.t.}\,\left(\forall\bmm{R}^{i}\right)\,\bm{q}^{\dagger}_{\perp}\bmm{G}_{ds}^{\circ\dagger}\bmm{R}^{i}\bm{t} = \bm{t}^{\dagger}\left(\bm{\chi}^{-1\dagger}-\bmm{G}_{dd}^{\circ\dagger}\right)\bmm{R}^{i}\bm{t}, 
    \nonumber \\
    &\lVert\bmm{Q}^{\dagger}\bm{q}_{\perp}\rVert_{2}^{2} = 0
    \text{\,\&\,}\lVert\bm{q}_{\perp}\rVert_{2}^{2} = 1,
    \label{eq:mainForm}
\end{align}
which is a QCQP over the combined vector $\left\{\bmm{q}_{\perp},\bmm{t}\right\}$. 
While such programs are generally difficult to solve\;\cite{park2017general}, bounds on $\sigma_{n}$ are nevertheless obtainable by Lagrangian duality or semi-definite relaxations\;\cite{molesky2020hierarchical,gustafsson2020upper,kuang2020computational}.
\subsection{Optimization Forms: $W$- and $P$-Operators}
\noindent
From bounds on $\sigma_{n}\left(\bmm{G}_{rs}\right)$, bounds on $\sigma_{n}$ for any other major scattering operators can be derived with relative ease, e.g. Eq.\eqref{eq:channelAmplitudeConversions}. 
For example, because the $\bmm{W}$-operator satisfies the operator relation $\left(\bmm{1}-\bmm{\chi}\bmm{G}^{\circ}\right)\bmm{W} = \bmm{1}$ and $\bmm{G} = \bmm{G}^{\circ}\bmm{W}$\;\cite{molesky2020t}, 
\begin{align}
    \bmm{\chi}\bmm{G} = \bmm{W} - \bmm{1}\Rightarrow 
    \bmm{W}_{rs} = \bmm{\chi}_{rr}\bmm{G}_{rs}
    \label{eq:gwRelation}.
\end{align}
Therefore, by the $1\,|\,n$ inequality, 
$$
    \sigma_{n}\left(\bmm{W}_{rs}\right)\leq\sigma_{1}\!\left(\bmm{\chi}_{rr}\right)\,\sigma_{n}\left(\bmm{G}_{rs}\right).
$$
Still, as argued in \emph{Formulation}, it is often useful to consider other operator descriptions. 
In particular, to simultaneously design the sender, mediator, and receiver of a fully coupled electromagnetic communication system (i.e. a system wherein scattering interactions between each of the subsystems substantially modify overall response characteristics), it is typically easier to work in terms of excited and received currents as defined by the $W$-operator, or ``field-power'' in terms of the $P$-operator. 
Via Eq.~\eqref{eq:WInvertedA} and the identity $\bmm{W}_{rr}^{\left(i\right)}\bm{\chi}_{rr} = \bmm{T}_{rr}^{\left(i\right)}$, consider the form
$$
    \bmm{W}_{sr}^{\dagger}\bmm{W}_{rs} = \bmm{W}_{ss}^{\dagger}\bmm{G}_{sr}^{\circ\dagger}\bmm{T}_{rr}^{\left(i\right)\dagger}\bmm{T}^{\left(i\right)}_{rr}\bmm{G}^{\circ}_{rs}\bmm{W}_{ss}.
$$
Suppose the receiver is decomposable into $R_{k}$ subregions wherein prescribed materials may be structured\;\cite{amaolo2024can}. 
Through the scattering relation of the $\bmm{T}^{\left(i\right)}_{rr}$-operator
$$
    \bmm{T}^{\left(i\right)\dagger}_{rr_{k}}=
    \bmm{T}^{\left(i\right)\dagger}_{rr_{k}}\left(\bm{\chi}^{-1}_{r_{k}r_{k}}-\bmm{G}^{\circ}_{r_{k}r}\right)\bmm{T}^{\left(i\right)}_{rr},
$$
where $\bm{\chi}^{-1}_{r_{k}r_{k}}$ may again be assumed to occupy its entire design volume, $\bmm{W}_{rs}^{\dagger}\bmm{W}_{rs}$ may be expanded as 
\begin{align}
    &\bmm{W}_{sr}^{\dagger}\bmm{W}_{rs} = -\sum_{r_{k}}\frac{\zeta_{r_{k}}}{2i}\bmm{W}_{ss}^{\dagger}\Bigg(
    \bmm{G}^{\circ\dagger}_{sr}\left(\bmm{T}_{rr_{k}}^{\left(i\right)\dagger}-\bmm{T}_{r_{k}r}^{\left(i\right)}\right)\bmm{G}^{\circ}_{rs}
    \nonumber \\
    &
    + \bmm{G}^{\circ\dagger}_{sr}\bmm{T}_{rr}^{\left(i\right)\dagger}\left(\bmm{G}^{\circ}_{r_{k}r}-\bmm{G}^{\circ\dagger}_{rr_{k}}\right)\bmm{T}_{rr}^{\left(i\right)}\bmm{G}^{\circ}_{rs}\Bigg)\bmm{W}_{ss}.
    \nonumber 
\end{align} 
Collecting terms and using the off-diagonal of Eq.\;\eqref{eq:WInvertedA}
\begin{align}
    \bmm{W}_{sr}^{\dagger}\bmm{W}_{rs} =
    \sum_{k} \zeta_{r_{k}}
    \bmm{W}^{\dagger}_{su}\left(-\bmm{G}^{\circ}_{r_{k}u}\right)^{\mathsf{a}}\bmm{W}_{us},
    \label{eq:wCFW}
\end{align}
with $\zeta_{r_{k}} = \left|\chi_{r_{k}}\right|^{2}/\Im\chi_{r_{k}}$.
Because $\bmm{G}^{vac\,\mathsf{a}}_{r_{k}r_{k}}$ is low rank (with eigenvectors representing the radiative states of the receiver volume\;\cite{molesky2019bounds}) and $\bmm{G}^{vac\,\mathsf{a}}_{rs}$ and $\bmm{G}^{vac\,\mathsf{a}}_{rm}$ are low rank by as a consequence of begin off-diagonal blocks of the vacuum Green operator, Eq.\;\eqref{eq:wCFW} is a low rank operator. 
To simplify further expressions we will collapse this sum to a single term going forward. 
The QCQP for $\sigma_{n}\left(\bmm{W}^{\dagger}_{sr}\bmm{W}_{rs}\right)$ given by the CFW theorem is then 
\begin{align}
    &\max_{\bm{q}_{\perp},\bm{w}}\,\zeta_{r}\,\bm{w}^{\dagger}\left(-\bmm{G}_{ru}^{\circ}\right)^{\mathsf{a}}\bm{w}~~\text{s.t.}
    \label{eq:mainTwo} \\
    &\left(\forall\bmm{R}^{i}\right)\,\left(\bm{w}-\bm{q}_{\perp}\right)^{\dagger}\bmm{R}^{i}\left(\bmm{1}-\bmm{\chi}\bmm{G}^{\circ}\right)\bm{w} = 
    \left(\bm{w}-\bm{q}_{\perp}\right)^{\dagger}\bmm{R}^{i}\bm{q}_{\perp},
    \nonumber \\
    &\lVert\bmm{Q}^{\dagger}\bm{q}_{\perp}\rVert_{2}^{2} = 0
    \text{\,\&\,}\lVert\bm{q}_{\perp}\rVert_{2}^{2} = 1,
    \nonumber
\end{align}
where $\bmm{\chi}$ is again the susceptibility operator of the structure having material occupying all design points, $\bm{w} = \bmm{W}\bm{q}_{\perp}$ is an unknown optimization vector confined to the combine receiver and mediator volume, $\bm{q}_{\perp}$ is an optimization vector confined to the source volume, and $\bmm{Q}$ is the matrix of the specified orthonormal bases $Q^{n-1}$. 
Given the computational cost associated with generic selections of $\bmm{Q}$ for larger $n$ values of Eq\;\eqref{eq:mainTwo}, we suggest that most implementations begin with the special basis choices described in \emph{Computational Expedients: System Design}.
\\ \\
As both $\bmm{P}_{rs}$ and $\bmm{W}_{rs}$ are restricted to the portion of the sender and receiver volumes occupied by material, the conversion 
$$
    \bmm{P}_{rs} = 2\,\chi_{rr}^{\mathsf{a}/2}\bmm{\chi}_{rr}^{-1}\bmm{W}_{rs}\bmm{\chi}^{\mathsf{a}/2}_{ss},
$$
reducing to Eq.\;\eqref{eq:channelAmplitudeConversions} in the case of isotropic electric media, together with Eq.\;\eqref{eq:mainTwo} is sufficient for determining bounds on $\sigma_{n}\left(\bmm{P}_{rs}\right)$ in most circumstances. 
For media with extremely low loss, it is plausible that inserting this relation into Eq.\;\eqref{eq:mainTwo} will improve numerical stability.
In particular, making the global definition $\bmm{P} = 2 \bm{\chi}^{\mathsf{a}/2}\bm{\chi}^{-1}\left(\bmm{W}-\bmm{1}\right)\bm{\chi}^{\mathsf{a}/2}$---with all $\bm{\chi}$ factors expanded to the fill the design regions, so that the complete structural information is encoded in $\bmm{W}-\bmm{1}$---the form of the general constraint for $\bmm{P}$ is 
\begin{align}
    2\,&\bmm{P}^{\dagger}\bmm{R}^{i}\bm{\chi}^{-\mathsf{a}/2}\bm{\chi}^{\dagger}\bmm{G}^{\circ}\bm{\chi}^{\mathsf{a}/2} = 
    \label{eq:PconstraintForm} \\
    &\bmm{P}^{\dagger}\bmm{R}^{i}\left(\bm{\chi}^{-\mathsf{a}/2}\chi^{\dagger}\bm{\chi}^{-\mathsf{a}/2}-\bm{\chi}^{-\mathsf{a}/2}\chi^{\dagger}\bmm{G}^{\circ}\bm{\chi}\bm{\chi}^{-\mathsf{a}/2}\right)\bmm{P}.
    \nonumber
\end{align} 
In the case of a single material Eq.\;\eqref{eq:PconstraintForm} simplifies to  
$$
    2\,\bmm{P}^{\dagger}\bmm{R}^{i}\bm{\chi}^{\dagger}\bmm{G}^{\circ} = 
    \bmm{P}^{\dagger}\bmm{R}^{i}\left(\bm{\chi}^{-\mathsf{a}/2}\chi^{\dagger}\bm{\chi}^{-\mathsf{a}/2}-\zeta\,\bmm{G}^{\circ}\right)\bmm{P}.
$$
which, via its anti-symmetric component, shows that the spectral norm of $\bmm{P}$ remains bounded in all circumstance.
\subsection{Subspace Optimization}
\noindent
Eqs.\;\eqref{eq:mainForm} and \eqref{eq:mainTwo} give bounds on $\sigma_{n}$ that depend on the choice of $Q^{n-1}$. 
As an extension, it is natural to consider optimizing this selection of $Q^{n-1}$ to obtain the tightest bound. 
\\ \\
Let $\bm{v}$ be the concatenated vector formed from $\bm{q}_{\perp}$ and $\bm{t}$ in the case of Eq.\;\eqref{eq:mainForm}, or $\bm{q}_{\perp}$ and $\bm{w}$ in the case of Eq.\;\eqref{eq:mainTwo}. 
Taking $\bmm{H}_{o}$ to be the Hermitian matrix of the objective, $\bmm{H}_{j}$ to be the Hermitian matrix of the $j^{th}$ scattering constraint, $\bmm{H}_{Q}$ to be the Hermitian matrix (or matrices) of the subspace orthogonality condition(s), and $\bmm{H}_{1}$ to be the matrix of the normalization constraint(s), the Lagrangians of programs defined by Eqs.\;\eqref{eq:mainForm} and Eqs.\;\eqref{eq:mainTwo} can be reformulated as  
\begin{equation}
    \mathcal{L}\left(\bm{v},\bmm{\mu},\mu_{Q},\bmm{H}_{Q}\right) 
    = \bm{v}^{\dagger}\left(\bm{\mu}\cdot\bmm{H} + \mu_{Q}\bmm{H}_{Q}\right)\bm{v} + \mu_{1},
    \label{eq:Lagrangian}
\end{equation}
with $\bm{\mu}\cdot\bmm{H} = \bmm{H}_{o} - \mu_{1}\bmm{H}_{1} - \sum_{j}\mu_{j}\bmm{H}_{j}$.
(Additional details on this transformation to purely quadratic forms are given in the supplementary information of Ref.\;\cite{kuang2020maximal} and Ref.\;\cite{molesky2025inferring}). 
\\ \\
The dual of Eq.\;\eqref{eq:Lagrangian}, upper bounding $\sigma_{n}^{2}$, is
\begin{align}
    &\mathcal{D}(\bmm{\mu},\mu_{Q},\bmm{H}_{Q})\equiv\max_{\bm{v}}\,\mathcal{L}\left(\bm{v},\bmm{\mu},\mu_{Q},\bmm{H}_{Q}\right)  
    \nonumber\\
    &= 
    \begin{cases} 
        \mu_{1} & \bm{\mu}\cdot\bmm{H} - \mu_{Q}\bmm{H}_{Q}\preceq 0 \\
        \infty & \text{otherwise} 
    \end{cases}.
    \label{eq:mainFormUpperBound}
\end{align}
$\bmm{H}_{Q} = \bmm{0}$ for $\sigma_{1}$, and an optimized subspace bound is given by minimizing the dual over the negative semi-definite cone $\bm{\mu}\cdot\bmm{H}\preceq 0$. 
Lower bounds on $\sigma_{1}$ can be similarly determined by beginning with the lower bound of Eq.\;\eqref{eq:CFWeigbounds} and following an analogous sequence of steps.
\\ \\
Minimization of the upper bound determined by Eq.\;\eqref{eq:mainFormUpperBound} for subsequent singular values is equated with the optimization program 
\begin{align}
    &\min_{\bmm{\mu},\mu_{Q},\bmm{H}_{Q}}\,\mu_{1}
    \label{eq:genSubspaceOpt} \\
    &\,\text{s.t.}\,\bm{\mu}\cdot\bmm{H} - \mu_{Q}\bmm{H}_{Q}\preceq 0,\,\bmm{H}_{Q}^{2} = \bmm{H}_{Q}\,\&\,\tr\left[\bmm{H}_{Q}\right] = k - 1.
    \nonumber
\end{align}
As $\bm{\mu}\cdot\bmm{H} - \mu_{Q}\bmm{H}_{Q}\preceq 0$ and $\bmm{H}_{Q}^{2} = \bmm{H}_{Q}$ are equivalent to 
\begin{align}
    &\left(\forall\bmm{N}_{k}\right)\,\tr{\left[\bmm{N}_{k}\left(\bmm{H}\left(\bmm{\mu}\right) - \mu_{Q}\bmm{H}_{Q}\right)\right]}\leq 0
    \nonumber\\ 
    &\left(\forall\bmm{K}_{ij}\right)\,\tr{\left[\bmm{H}_{Q}\bmm{K}_{ij}\bmm{H}_{Q}\right]} = \tr{\left[\bmm{K}_{ij}\bmm{H}_{Q}\right]},
    \label{eq:subOptMatConstraints}
\end{align}
with $\bmm{N}_{k}$ a positive semi-definite rank-one matrix with unit eigenvalue, $\bmm{K}_{ij}$ standing for $\left(\bmm{\delta}_{i}\otimes\bmm{\delta}_{j}+ \bmm{\delta}_{j}\otimes\bmm{\delta}_{i}\right)/2$, and $\bmm{\delta}_{j}$ a vector having value one at position $j$ and zero at all other indices, Eq.\;\eqref{eq:genSubspaceOpt} is again a QCQP. 
Any feasible point of Eq.\;\eqref{eq:genSubspaceOpt} yields a bound on $\sigma_{n}$. 
\subsection{Singular Value Bounds and Density of States Divergence}
\noindent
The implication of Eq.\;\eqref{eq:anaResultPrsAmp} that $\left(\forall k\right)\,\sigma_{k}\left(\bmm{G}_{rs}\right)\leq\left(4\Im\left(\chi_{s}\right)\Im\left(\chi_{r}\right)\right)^{-1}$ for arbitrarily structured media has the potential to cause confusion in light of two well known results concerning Green functions: (1) even in vacuum, the Green function possesses unbounded (singular) components; (2) the spectral power drawn from a point dipole (the local density of states) diverges in the limit of vanishing separation with a lossy medium. 
Brief explanations as to why these facts are not contradictory are provided below.  
\\ \\
The first tension, between bounded singular values and unbounded singular components, is entirely semantic. 
By a Green operator, we mean the object that inverts a particular electromagnetic wave equation, and not the associated integral kernel, which we refer to as the Green function, see \emph{Formulation}. 
Under this operator definition, singular components always act in conjunction with integration (convolution) to produce well defined linear mappings, and it these mappings that the bounds refer to. 
\\ \\
Resolution of how a point dipole (a vector) creates unbounded power transfer, when all singular values of $\bmm{G}_{rs}$ are bounded, is also largely a matter of definitions. 
Namely, a point dipole is an idealized limit of a sequence of shrinking spatial extent (or increasing concentration) subject to an $L^{1}$-norm condition that sets the fixes the total current: e.g. 
\begin{align*}
    \lim_{n\rightarrow\infty}\bm{j}_{n} = \sqrt{\frac{n}{2\pi}}^{\,3}\exp\left(-n r^{2}/2\right)\hat{\bm{z}}~~~\left(\forall n\,\,\int_{V}\!dV\,\bm{j}_{n} = \hat{\bm{z}}\right).
\end{align*}
The $L^{2}$-norm of such a sequence, which is the relevant norm for determining singular values, is always unbounded: for the example given above, 
$$
    \left(\int_{V}dV\,\bm{j}^{*}_{n}\,\bm{j}_{n}\right)^{1/2} = \left(\frac{n}{4\pi}\right)^{3/4}.
$$
As such, for the same reason that the total energy of a dipole is undefined\;\cite{chao2023maximum}, a single dipole, despite its vector status, can access an unbounded number of channels. 
\\ \\
Although the physical end result is the same, $L^{1}$-normalization is not the cause of the divergence observed in radiative heat transfer between bodies being brought into contact (in local response theory);
the fluctuation-dissipation theorem asserts that correlations of stochastic currents are proportional to the asymmetric component of the associated susceptibility, which fixes the $L^{2}$-norm.
Rather, near-field radiative heat transfer divergence occurs because classical electromagnetics places no feature size limit on current densities. 
The number of orthogonal current excitations that can exist with any material object of finite size is unbounded, which (in the limit of vanishing separation) leads to heat flux through an unbounded number of channels.  
\section{\S Computational Expedients}
\noindent
The programs proposed by Eqs.\;\eqref{eq:mainForm} and \eqref{eq:mainTwo}, and to an even greater extent Eq.\;\eqref{eq:subOptMatConstraints}, are computationally demanding for three-dimensional systems (at present boarding on impractical).  
Below, we describe how a specific basis choices can mitigate these costs, allowing singular value bounds up to $n\approx 100$ to be calculated (in parallel) for system design volumes approaching one hundred cubic wavelengths. 
\subsection{Mediator Design}
\label{sec:gChain}
\noindent
As the simpler of the two cases, we begin with the problem of designing a mediator to maximize the singular values of a sender-receiver Green operator $\bmm{G}_{rs}$.  
Throughout this subsection, the first $n$ right singular vectors of $\bmm{G}^{\circ}_{xy}$ will be denoted as $U^{n}_{xy}$ (total basis $U_{xy}$). 
The first $n$ left singular vectors of $\bmm{G}^{\circ}_{xy}$ will be similarly denoted as $V^{n}_{xy}$ (total basis $V_{xy}$). 
\\ \\
As an immediate application of Eq.\;\eqref{eq:CFWeigbounds} of the main text to Eq.\;\eqref{eq:greenScattering}, the definition of max implies
\begin{align}
    \sigma_{n}\left(\bmm{G}_{rs}\right) \leq
    &\max_{\bm{u}_{\perp}\in U^{n-1}_{ds\,\perp}}\bigg(\max_{\bm{v}^{a}\in V_{rs}}\Re\left(\bm{v}^{a\dagger}\bmm{G}_{rs}^{\circ}\bm{u}_{\perp}\right) +
    \nonumber \\
    &~\,\max_{\bm{v}^{b}\in V_{rd}}\Re\left(\bm{v}^{b\dagger}\bmm{G}^{\circ}_{rd}\bmm{T}_{d}\bmm{G}^{\circ}_{ds}\bm{u}_{\perp}\right)\bigg). 
    \label{eq:maxMaxForm}
\end{align}
To become acquainted some of the tradeoffs implicit in this from, note that by the global real power conservation
\begin{equation}
    \bmm{T}^{\mathtt{a}}_{d} = \bmm{T}^{\dagger}_{d}\left(\bmm{G}^{\circ}_{dd}-\bm{\chi}^{-1}\right)^{\mathtt{a}}\bmm{T}_{d}.
    \label{eq:gloRelPow}
\end{equation}
Consequently, the magnitude of any image vector produced by $\bmm{T}_{d}$ depends on the relative alignment of the image to the source: setting $\bm{t} = \bmm{T}_{d}\bm{s}$ and $\zeta_{d} = \lVert\chi_{d}\rVert^{2}_{2}/\Im\chi_{d}$ Eq.\;\eqref{eq:gloRelPow} implies that $\lVert\bm{t}\rVert_{2}\leq\zeta_{d}\,\Im\left(\bm{s}^{\dagger}\bm{t}\right)$\;\cite{miller2016fundamental}.
To maximize the bound on $\gamma_{rs\,n}$ stated by \eqref{eq:maxMaxForm}, $\bmm{T}_{d}$ should naively convert any $\bm{v}_{ds}^{k}$ into a vector lying largely in the subspace defined by the largest right eigenvectors of $\bmm{G}^{\circ}_{rd}$.  
If $\bmm{v}^{ds}_{k}$ does not overlap meaningfully with this subspace, then the constraint on the magnitude of the image may make this choice sub-optimal. 
\\ \\
Starting with the unrestricted square of the second term, 
$$
    \max_{\bm{u}\in U_{ds}}\bm{u}^{\dagger}\bmm{G}^{\circ\dagger}_{sd}\bmm{T}_{d}^{\dagger}\bmm{G}^{\circ\dagger}_{dr}\bmm{G}^{\circ}_{rd}\bmm{T}_{d}\bmm{G}^{\circ}_{ds}\bm{u},
$$
take $\bm{s}$ to be the coefficients of expansion of $\bm{u}\in U_{ds}$, and $\bm{t}_{k}$ to be the coefficients of expansion of $\bmm{T}_{d}\bm{v}^{k}_{ds}$ in $U_{rd}$. 
Substitution into Eq.\eqref{eq:maxMaxForm} then gives
\begin{equation}
    \max_{\bm{s},\bm{t}_{n},\bm{t}_{n+1},\ldots} \sum_{k}\left|s_{k}\right|^{2}\sigma_{k}^{2}\left(\bmm{G}^{\circ}_{ds}\right)\,\bm{t}_{k}^{\dagger}\bmm{D}^{\circ}_{dd}\bm{t}_{k},
    \nonumber
\end{equation}
with $\bmm{D}^{\circ}_{dd}$ the diagonal matrix of the singular value decomposition of $\bmm{G}^{vac\,\dagger}_{dr}\bmm{G}^{\circ}_{rd}$, subject to scattering constraints of the form specified by Eq.\;\eqref{eq:quadCstrts} for each $\bm{t}_{k}$.
If scattering constraints are only enforced on each $\bm{t}_{k}$ individually, then the computation of a chain singular value breaks down into a set of independent optimizations\,\footnote{Even though each $s_{k}$ represents an orthogonal excitation of the design, physical consistency demands that there should also be cross-constraints associated with the fact that each of these responses must be realized by a single structure\;\cite{molesky2021comm}.}: some $\sigma_{k}^{2}\left(\bmm{G}^{\circ}_{ds}\right)\,\bm{t}_{k}^{\dagger}\bmm{D}^{\circ}_{dd}\bm{t}_{k}$ term will be maximal, and the optimal choice of $\bm{s}$ is simply to set $\left|s_{k}\right|^{2} = 1$. 
To compute indexed channel amplitude bounds, the corresponding maximal singular vector of $\bmm{G}^{\circ}_{rd}$ is then excised (by rank removal), and a bound on the subsequent singular value of $\bmm{G}^{\circ\dagger}_{sd}\bmm{T}_{d}^{\dagger}\bmm{G}^{\circ\dagger}_{dr}\bmm{G}^{\circ}_{rd}\bmm{T}_{d}\bmm{G}^{\circ}_{ds}$ is given by the next best value of $\sigma_{k}^{2}\left(\bmm{G}^{\circ}_{ds}\right)\,\bm{t}_{k}^{\dagger}\bmm{D}^{\circ}_{dd}\bm{t}_{k}$. 
\\ \\ 
Determining bounds on each of the $\sigma_{k}^{2}\left(\bmm{G}^{\circ}_{ds}\right)\,\bm{t}_{k}^{\dagger}\bmm{D}^{\circ}_{dd}\bm{t}_{k}$ terms results in a list of QCQP sub-problems of a form similar to Eq.\;\eqref{eq:mainForm}:  
\begin{align}
    &\xi^{2}_{k} = \max_{\bm{t}_{k}\in\mathbb{C}^{n}}\bm{t}^{\dagger}_{k}\bmm{D}^{\circ}_{dd}\bm{t}_{k}
    \label{eq:tSingOpt} \\
    &\text{s.t.}\,\left(\forall\bmm{R}^{i}\right)\,\bm{d}_{k}^{\dagger}\bmm{R}^{i}\bm{t}_{k}
    = \bm{c}^{\dagger}_{k}\bmm{U}_{dr}^{\dagger}\left(\bm{\chi}^{-1}-\bmm{G}^{\circ}_{dd}\right)^{\dagger}\bmm{U}_{rd}\bmm{R}^{i}\bm{c}_{k} 
    \nonumber
\end{align}
where $\bm{d}_{k}$ is the vector of expansion coefficients of $\bm{v}_{ds}^{k}$ in the basis of $U_{rd}$.
While Eq.\;\eqref{eq:tSingOpt} remains generally difficult to solve, the method employed in Ref.\;\cite{chao2023maximum} shows that when the considered $\bmm{R}^{i}$ operators are limited to global real and reactive power conservation---$\bmm{R}^{\left(1\right)} = \bmm{1}$ and $\bmm{R}^{\left(2\right)} = i\bmm{1}$---limits can be efficiently computed for domains surpassing $100\,\lambda^{3}$ by Lagrange duality. 
Taking an additional step, if the constraints of Eq.\;\eqref{eq:tSingOpt} are limited to only real power conservation the solution of Eq.\;\eqref{eq:tSingOpt} is given by the semi-analytic form
\begin{equation}
    \bm{t}^{\star}_{k} = \frac{-i\alpha}{2}\left[\bmm{D}^{\circ}_{dd}+\alpha\bmm{U}_{dr}^\dagger\left(\bm{\chi}^{-1}\!\!-\bmm{G}^{\circ}_{dd}\right)^{\mathtt{a}}\bmm{U}_{rd}\right]_{\mathtt{p}}^{-1}\!\!\bm{d}_{k},
    \label{eq:ckstar}
\end{equation}
with $\left[\ldots\right]_{\mathtt{p}}^{-1}$ denoting the pseudo inverse of the enclosed operator, and $\alpha > 0$ determined by the location of the last zero of the function
\begin{equation}
    f\left(\alpha\right)\!= \bm{t}^{\star\dagger}_{k}\!\left(2\bmm{D}^{\circ}_{dd}+\alpha\bmm{U}_{dr}^\dagger\left(\bm{\chi}^{-1}\!\!-\bmm{G}^{\circ}_{dd}\right)^{\mathtt{a}}\bmm{U}_{rd}\right)\!\bm{t}^{\star}_{k}.
    \label{eq:alphaSolve}
\end{equation}
Because the matrices appearing in this function definition are positive semi-definite and negative semi-definite respectively, the value of $\alpha$ may be obtained by essentially any root solving method (\emph{Numerical Methods}).
For rapid evaluation, or as a solution sanity check, it is also possible to resort to the large $\alpha$ limit
$$
    \bm{t}_{k\star} = -i\left[\bmm{U}_{dr}^\dagger\left(\bm{\chi}^{-1}-\bmm{G}^{\circ}_{dd}\right)^{\mathsf{a}}\bmm{U}_{rd}\right]^{-1}\bm{d}_{k}/2,
$$
yielding the upper bound $\xi^{2}_{k}\leq\bm{v}^{k\dagger}_{sd}\bmm{Z}_{dd}^{\dagger}\bmm{D}^{\circ}_{dd}\bmm{Z}_{dd}\bmm{v}^{k}_{ds}/4$, with $\bmm{Z}_{dd} = \left[\left(\bm{\chi}^{-1}-\bmm{G}^{\circ}_{dd}\right)^{\mathsf{a}}\right]^{-1}$, in basis independent form.
Making the final simplification of discounting radiative loss, i.e. enforcing only the passivity inequality $\Re\left(i\bm{t}_{k}^{\dagger}\bm{d}_{k}\right) - \bm{t}^{\dagger}_{k}\left(i\bm{\chi}^{-1}\right)^{\mathtt{a}}\bm{t}_{k}\geq 0$, this last approximation simplifies to 
$$
    \xi_{k}^{2}\leq \zeta_{d}^{2}\sum_{j}\left| d_{k\,j}\right|^{2}\sigma_{j}^{2}\left(\bmm{G}^{\circ}_{rd}\right)/4.
$$
If the $U_{rd}$ coefficients of $\bm{d}_{k}$ are tightly centred around $j = k$, which may reasonably be expected based on reciprocity in high symmetry cases, then this reduction essentially reproduces the ``basis chaining'' form that maximizes the overall Frobenius norm. 
It is also interesting to note that because $\left\{\bmm{v}_{ds}^{k}\right\}$ is a basis for the vector space of fields in the design volume, even this simple estimate prevents fictitious over coupling to the largest singular values of $\bmm{G}^{\circ}_{ds}$.
Still, particularly when $\bmm{\chi}^{\mathsf{a}}$ is small or accurate values are important, more exact version of Eq.\;\eqref{eq:tSingOpt} should be favoured. 
\\ \\
Returning to the original objective, the preceding argument proves that 
\begin{align*}
    \max_{\bmm{u}_{\perp}\in U^{n-1}_{\perp}}&\max_{\bm{v}_{\perp}^{b}\in V_{rd}}\Re\left(\bm{v}_{\perp}^{b\dagger}\bmm{G}^{\circ}_{rd}\bmm{T}_{d}\bmm{G}^{\circ}_{ds}\bmm{u}_{\perp}^{b}\right)\leq
    \\
    &~\,\max_{k > n}\left\{\xi_{k}\,\sigma_{k}\left(\bmm{G}^{\circ}_{rd}\right)\right\}.    
\end{align*}
Decay in $\sigma_{k}\left(\bmm{G}^{\circ}_{rd}\right)$, and the ability to bound larger values of $k$ with more approximate versions of Eq.\;\eqref{eq:tSingOpt}, will typically limit the number of entries that must be practically considered in the max set, especially as $n$ grows large. 
$\left(\forall k\right)\,\xi_{k}\leq\zeta_{d}\,\sigma_{1}\left(\bmm{G}^{\circ}_{rd}\right)/2$, and so $k\geq p$ implies that $\xi_{k}\,\sigma_{k}\left(\bmm{G}^{\circ}_{ds}\right)\leq\zeta_{d}\,\sigma_{1}\left(\bmm{G}^{\circ}_{rd}\right)\sigma_{p}\left(\bmm{G}^{\circ}_{ds}\right)/2$. 
To guarantee that the max set is sufficiently large, so long as some separation exists between either the design and the source and the design and the receiver, $p$ may be found such that this bound is smaller than some existing element in the set\;\footnote{This estimate can be improved by tracking what portion of $U_{rd}$ has already been covered, and then optimizing $\bmm{d}_{k}$ over the remaining subspace.}.
\\ \\
As all structuring effects are confined to the already been treated scattering term
$$
    \tilde{\gamma}^{\circ}_{rs\,n} = \max_{\bm{u}_{\perp}\in U^{n-1}_{ds\,\perp}}\max_{\bm{v}_{\perp}^{a}\in V_{rs}}\Re\left(\bm{v}_{\perp}^{a\dagger}\bmm{G}_{rs}^{\circ}\bm{u}_{\perp}^{a}\right)
$$ 
will usually determine a contribution much smaller than $\max_{k > n}\left\{\xi_{k}\,\sigma_{k}\left(\bmm{G}^{\circ}_{ds}\right)\right\}$. 
If this is not the case, then right singular vectors of $\bmm{G}^{\circ}_{rs}$ can be considered in conjunction with $U^{n-1}_{rd}$ basis of $\bmm{G}^{\circ}_{ds}$ to form a more accurate subspace selection (e.g. guaranteeing that the contribution of first never surpasses that of the second). 
\\ \\
Since the roles of the target and source spaces in defining the smallest singular value may be freely exchanged, a final bound on $\gamma_{rs\,n}$ is given by
\begin{align}
    \sigma_{n}\left(\bmm{G}_{rs}\right)&\leq\min
    \begin{Bmatrix}
        \tilde{\gamma}^{\circ}_{rs\,n} + \xi_{n\star}\,\sigma_{n\star}\left(\bmm{G}^{\circ}_{ds}\right)\\
        \breve{\gamma}^{\circ}_{rs\,n} + \breve{\xi}_{n\star}\,\sigma_{n\star}\left(\bmm{G}^{\circ}_{rd}\right)
    \end{Bmatrix},
    \label{eq:dirMedOnlyBnd}
\end{align}
with a $\breve{}$ accent indicating that $U^{n-1}_{ds\,\perp}$ has been exchanged for $V^{n-1}_{rd\,\perp}$: $\breve{\xi}_{n\star}$ being defined by substituting $\bmm{V}_{ds}$ for the basis of $\bmm{U}_{rd}$ in Eq.\;\eqref{eq:tSingOpt}, and $n\star$ denotes the optimal $k$ of $\max_{k>n}\left\{\ldots\right\}$.
\subsection{System Design}
\label{sec:wChain}
\noindent
If the receiver, source, and mediator are all supposed to be freely structurable, the singular value decompositions of $\bmm{G}_{rd}^{\circ}$ and $\bmm{G}_{ds}^{\circ}$ enabling the mediator design approach described in the previous subsection will contain a large number (proportional to the computational discretization) of near unit strength entries that are almost completely confined to the receiver and source. 
We have consistently found that these terms do not actually contribute to indexed singular value bounds for practical values of $n$.
Yet, we do not have a simple and rigorous method to remove them a priori. 
As such, neither $\sigma_{n}\left(\bmm{G}^{\circ}_{ds}\right)$ nor $\sigma_{n}\left(\bmm{G}^{\circ}_{rd}\right)$ natively exhibits decay in such settings, adding a non-trivial complexity to determining which vectors to use in Eq.\;\eqref{eq:dirMedOnlyBnd}. 
Rather than working through this difficulty, we believe that it is technically simpler to use Eq.\;\eqref{eq:wCFW} as described below. 
\\ \\
As a consequence of Eq.\;\eqref{eq:wCFW} and $\bmm{W}\bm{\chi} = \bmm{T}$, 
$$
    \bmm{T}_{ss}^{\dagger}\bmm{X}_{ss}\bmm{T}_{ss} = \bmm{T}^{\dagger}_{su}\left(-\bmm{G}^{\circ}_{ru}\right)^{\mathsf{a}}\bmm{T}_{us},
$$
with $\bmm{X}_{ss}$ as in \eqref{eq:oldBoundsFrobenius}. 
By use of Eq.\;\ref{eq:mainTwo}, rank removal, and $\bmm{W}\bm{\chi} = \bmm{T}$, bounds on $\sigma_{n}\left(\bmm{P}_{rs}\right)^{2}$ are then given by 
\begin{align}
    &\max_{\bm{x}\ni\lVert\bm{x}\rVert^{2}_{2} = 1}\bigg(
    \frac{4}{\zeta_{s}}\,\bm{x}^{\dagger}\bmm{T}_{su}^{\dagger}\left(-\bmm{G}^{\circ}_{ru}\right)^{\mathsf{a}\,\left(n\right)}\bmm{T}_{us}\bmm{x}
    \nonumber \\
    &~\text{s.t.}\,\bmm{T}_{ss}^{\mathsf{a}} = 
    \bmm{T}_{su}^{\dagger}
    \left(\bm{\zeta}_{ss}^{-1} + \left(-\bmm{G}^{\circ}_{ru}\right)^{\mathsf{a}} + \left(\bmm{G}_{ss}\right)^{\mathsf{a}}\right)\bmm{T}_{us},
    \nonumber \\
    &~~~~~\,\bmm{T}_{su}^{\dagger}\left(-\bmm{G}^{\circ}_{ru}\right)^{\mathsf{a}}\bmm{T}_{us} = \bmm{T}_{sr}^{\dagger}\bmm{\zeta}_{rr}^{-1}\bmm{T}_{rs},
    \nonumber \\
    &~\&\,\left(\forall\bmm{R}^{i}\right)\bmm{R}^{i}\bmm{T} = \bmm{T}^{\dagger}\left(\bm{\chi}^{-1}_{uu} - \bmm{G}^{\circ}_{uu}\right)^{\dagger}\bmm{R}^{i}\bmm{T}\bigg),
    \label{eq:wChainProg}
\end{align}
with all constraints applicable to any vector in the system. 
Through the eigenbases of various Green operator, several versions of Eq.\;\eqref{eq:wChainProg} can be solved numerically. 
For the purposes of treating limit sizes and extracting semi-analytic understanding, below we will continue with additional simplifications.  
\\ \\
Because $\bm{x}$ is properly confined to reside in the source region, the first constraint is equivalently rewritten as  
$$
    \bm{x}^{\dagger}\bmm{T}_{uu}^{\mathsf{a}}\bm{x} = 
    \bm{x}^{\dagger}\bmm{T}_{uu}^{\dagger}
    \left(\bm{\zeta}_{ss}^{-1} + \left(-\bmm{G}^{\circ}_{ru}\right)^{\mathsf{a}} + \left(\bmm{G}_{ss}^{\circ}\right)^{\mathsf{a}}\right)\bmm{T}_{uu}\bm{x}.
$$
Adding the total real power conservation constraint 
$$
     \bm{x}^{\dagger}\bmm{T}_{uu}^{\mathsf{a}}\bm{x}=
     \bm{x}^{\dagger}\bmm{T}_{uu}^{\dagger}\left(\bmm{\zeta}^{-1}_{uu}
     + \left(\bmm{G}_{uu}^{\circ}\right)^{\mathsf{a}}\right)\bmm{T}_{uu}\bm{x},
$$
the above relation is equivalent to 
\begin{align*}
    2\,\bm{x}^{\dagger}\bmm{T}_{uu}^{\mathsf{a}}\bm{x} =\, &
    \bm{x}^{\dagger}\bmm{T}_{uu}^{\dagger}
    \left(\bm{\zeta}_{uu}^{-1} + \left(-\bmm{G}^{\circ}_{ru}\right)^{\mathsf{a}} + 
    \left(\bmm{G}_{uu}^{\circ}\right)^{\mathsf{a}}\right)\bmm{T}_{uu}\bm{x}\,+ \\
    &\bm{x}^{\dagger}\bmm{T}_{uu}^{\dagger}
    \left(\bm{\zeta}_{ss}^{-1} + \left(\bmm{G}_{ss}^{\circ}\right)^{\mathsf{a}}\right)\bmm{T}_{uu}\bm{x}
\end{align*}
Consequently---as if so desired $\bm{x}$ can always be selected entirely within the source to reproduced the physical inequality---$\bm{x}$ can be extended to the full system to yield the relaxation of Eq.\;\eqref{eq:wChainProg}
\begin{align}
    &\sigma_{n}^{2}\left(\bmm{P}_{rs}\right)\leq\frac{4}{\zeta_{s}}\max_{\bm{x}\ni\lVert\bm{x}\rVert^{2}_{2} = 1}\bigg(
    \,\bm{t}^{\dagger}\left(-\bmm{G}^{\circ}_{ru}\right)^{\mathsf{a}\,\left(n\right)}\bm{t}
    \label{eq:simpleWChainProg} \\
    &\text{s.t.}\,2\,\Im\left\{\bm{x}^{\dagger}\bm{t}\right\} =
    \bm{t}^{\dagger}
    \left(\bm{\zeta}_{uu}^{-1} + \left(-\bmm{G}^{\circ}_{ru}\right)^{\mathsf{a}} + \left(\bmm{G}_{uu}^{\circ}\right)^{\mathsf{a}}\right)\bm{t}\, +
    \nonumber \\
    &~~~~~~~~~~~~~~~~~~~~~\,\bm{t}^{\dagger}
    \left(\bm{\zeta}_{ss}^{-1} + \left(\bmm{G}_{ss}^{\circ}\right)^{\mathsf{a}}\right)\bm{t}
    \bigg),
    \nonumber
\end{align}
with $\bm{t} = \bmm{T}_{uu}\bm{x}$.
Supposing $\bm{x}$ and $\bm{t}$ to be independent variables, it is clear that any solution of Eq.\;\eqref{eq:simpleWChainProg} will be characterized by $\bm{x}$ being a unit vector along $\bm{t}$ with relative phase of $-i$.
Letting $t$ be the magnitude of $\bm{t}$, and setting $\bm{t}^{\dagger}\left(-\bmm{G}^{\circ}_{ru}\right)^{\mathsf{a}\,\left(n\right)}\bm{t} = t^{2}\xi^{\left(n\right)}$, $\bm{t}^{\dagger}\left(-\bmm{G}^{\circ}_{ru}\right)^{\mathsf{a}}\bm{t} = t^{2}\xi$, $\bm{t}^{\dagger}\left(\bm{\zeta}_{ss}^{-1} + \left(\bmm{G}_{ss}^{\circ}\right)^{\mathsf{a}}\right)\bm{t} = t^{2}\alpha$, and $\bm{t}\left(\bmm{G}_{uu}^{\circ}\right)^{\mathsf{a}}\bm{t}^{2} = \rho$, Eq.\;\eqref{eq:simpleWChainProg} becomes 
$$
    \sigma_{n}^{2}\left(\bmm{P}_{rs}\right)\leq\frac{4}{\zeta}\max_{\bm{t}}~t^{2}\xi^{\left(n\right)}~~
    \text{s.t.}\,2 =
    t
    \left(\bm{\zeta}^{-1} + \xi + \rho + \alpha\right),
    \nonumber
$$
where for simplicity it has been assumed that $\bmm{\zeta}^{-1}_{uu} = \zeta^{-1}\bmm{1}_{uu}$. 
The solution of this optimization, in terms of yet undetermined parameters $\xi^{\left(n\right)}$, $\xi$, $\rho$ and $\alpha$, is 
$$
    \sigma_{n}^{2}\left(\bmm{P}_{rs}\right)\leq
    16\,\zeta\,\xi^{\left(n\right)} / \left(1 + \zeta\left(\xi + \rho + \alpha\right)\right)^{2}.
$$
As $\xi\geq\xi^{\left(n\right)}$ by definition, $\xi^{\left(n\right)}\leq\sigma_{n}\left(\left(-\bmm{G}^{\circ}_{ru}\right)^{\mathsf{a}}_{+}\right)$---with $\left(-\bmm{G}^{\circ}_{ru}\right)^{\mathsf{a}}_{+}$ the restriction of $\left(-\bmm{G}^{\circ}_{ru}\right)^{\mathsf{a}}$ to the subspace spanned by its eigenvectors with positive eigenvalues---and $\alpha\geq 1 - \zeta\xi$ by equivalent descriptions of power transfer, the selection of (super) optimal coefficients for this form, together with Eq.\;\eqref{eq:anaResultPrsAmp},
gives
\begin{equation}
    \sigma_{n}^{2}\left(\bmm{P}_{rs}\right)\leq\min\left\{1,\,4\,\zeta\,\sigma_{n}\!\left(\left(-\bmm{G}^{\circ}_{ru}\right)^{\mathsf{a}}_{+}\right)\right\}.
    \label{eq:asymGurAnalytic}
\end{equation}
Because $\left(-\bmm{G}^{\circ}_{ru}\right)^{\mathsf{a}}_{+}$ subsumes contributions from $-\left(\bmm{G}^{\circ}_{rr}\right)^{\mathsf{a}}$, which act to discount reradiation from the receiver in the determination of how much power has been transferred from the source, $\zeta\,\sigma_{n}\!\left(\left(-\bmm{G}^{\circ}_{ru}\right)^{\mathsf{a}}_{+}\right)$ is, in some instances, substantially smaller than $\zeta^{2}\,\sigma_{n}^{2}\left(\bmm{G}^{\circ}_{rs}\right)$.
The consequences of this difference has been found to be particularly relevant for wavelength scale ($\lesssim\lambda^{3}$) senders and receivers separated by subwavelength distances.
\section{\S Numerical Methods}
\noindent
All the code used to generate the 3D bounds calculations along with the verifications of our numerical methods is available at \url{https://github.com/PaulVirally/Photonic-System-Channels}.
\subsection{Discretization and Matrix-Free Green Operators}
\noindent
All vacuum Green operators are computed with an in-house volume-integral equation implementation of the electric Green tensor on uniform Cartesian grids (available through the Julia package \texttt{GilaElectromagnetics.jl}).
The kernel is discretized in a Galerkin fashion over piecewise-constant basis functions, with the weak singularity of the self-interaction handled analytically through a singularity subtraction method.
By choosing open boundary conditions, the discretized operator is block-Toeplitz due to translational invariance.
A circulant embedding then allows matrix-vector products to be evaluated by fast Fourier transforms in $\mathcal{O}\left(N\log N\right)$ time and $\mathcal{O}\left(N\right)$ memory, with $N = 3n_{x}n_{y}n_{z}$ the number of field degrees of freedom. 
The Green operators are never explicitly formed; rather, all uses of $\bmm{G}^{\circ}$ are through this fast application\;\cite{polimeridis2014computation}.
\subsection{Randomized Singular Value Decomposition}
\label{sec:rsvd}
\noindent
Performing a singular value decomposition of the vacuum operators $\bmm{G}^{\circ}_{xy}$ for the different regions $xy \in \set{rs, rd, dd, ds}$ is, in general, a non-trivial task due to the size associated with the discretized integral operators.
For a three dimensional problem discretized on a cuboid grid of $n_{i}$ cells in each direction, $i \in \set{x, y, z}$, the size of the operator is $(3n_{x}n_{y}n_{z})\times(3n_{x}n_{y}n_{z})$.
This $\mathcal{O}\left(n^{6}\right)$ scaling in memory precludes explicit storage and effectively requires matrix-free methods for any reasonably large system.
For the examples presented in the main text, we have used randomized singular value decomposition (RSVD) following Algorithm 4.4 of \cite{halko2011finding} with power iterations and oversampling (available through the Julia package \texttt{MatrixFreeRandomizedLinearAlgebra.jl}).
\\ \\
Specifically, to return $k$ singular components, $t = k+p$ test vectors are drawn ($p=k$ for the results in Fig.\ \ref{fig:svdBounds} and $p=50$ for the results in Fig.\ \ref{fig:traceBounds}) following $q$ power iterations (here, $q=14$).
The method constructs a small matrix $\bmm{Q}_{xy}\ni\bmm{Q}_{xy}\bmm{Q}^\dagger_{yx}\bmm{G}^{\circ}_{xy}\approx\bmm{G}^{\circ}_{xy}$ and decomposes $\bmm{G}^{\circ}_{xy}\approx\bmm{V}_{xy}\bmm{D}_{xy}\bmm{U}^\dagger_{xy} = \bmm{Q}_{xy}\tilde{\bmm{V}}_{xy}\bmm{D}_{xy}\bmm{U}_{yx}^\dagger$ where $\bmm{Q}_{yx}^\dagger \bmm{G}^{\circ}_{xy} = \tilde{\bmm{V}}_{xy}\bmm{D}_{xy}\bmm{U}_{yx}^\dagger$ is the standard singular value decomposition of a tall-and-skinny matrix which can be done efficiently on a dense matrix. 
The matrix with orthonormal columns, $\bmm{Q}_{xy}$, turns out to be easily computed by orthonormalizing the matrix $\bmm{Y}_{xy} = \left(\bmm{G}^{\circ}_{xy}\bmm{G}^{\circ\dagger}_{yx}\right)^q\bmm{G}^{\circ}_{xy}\bmm{\Omega}$ where $\bmm{\Omega}$ is a tall-and-skinny matrix with $t$ columns whose entries are drawn from a standard Gaussian.
The use of $q\geq 0$ power iterations sharpens the singular value spectrum.
Intuitively, random test vectors probe all directions; after application by $\mathbf G^{\circ}_{xy}$ (and power iterations), their images concentrate along the dominant column space, so $\mathbf Q_{xy}$ captures the top left singular subspace with high probability (oversampling provides a safety margin \cite{halko2011finding}).
\\ \\
The particular choice of $q = 14$ and $p \in\left\{50, k\right\}$ used here was made by ensuring the relative RSVD dual residuals, $\lVert\bmm{G}^{\circ\dagger}\bmm{v}_{i} - \sigma_i\bmm{u}_{i}\rVert_2 / \sigma_{i}$ were smaller than $10^{-12}$ and, for the mediator chains, by comparing the largest singular value of $\bmm{G}^{\circ}_{rs}\bmm{U}^\perp_{n-1}$ and $\bmm{V}^\perp_{n-1}\bmm{G}^{\circ}_{rs}$ against $\sigma_{n}\left(\bmm{G}^{\circ}_{rs}\right)$ for all $n\leq k$.
The target rank $k$ varies per system, but was chosen to saturate on-node memory.
For instance, in the $4\lambda\times 4\lambda\times 4 \lambda$ mediator result, a rank of $k=57$ targeted to use up the 80\, GB of VRAM available on an NVIDIA H100. Note that at the discretization level used of 32\,768 voxels per cubic wavelength, the associated operator would have an actual rank of 3\,145\,728.
\subsection{Zeros of $f(\alpha)$}\label{subsec:zeroFinder}
\noindent
Via algebraic manipulations a numerically efficient and stable expression for Eq.\;\eqref{eq:alphaSolve} can be obtained as follows.
Let $\bmm{A} = \bmm{U}_{dr}^\dagger \left(\bmm{\chi}^{-1\dagger} - \bmm{G}^{\circ\dagger}_{dd}\right)^\mathtt{a}\bmm{U}_{rd}$. As $\bmm{A}$ is Hermitian, it is unitarily diagonalizable: $\bmm{A} = \bmm{Q}\bmm{\Lambda}\bmm{Q}^\dagger$.
As such, by factorizing, Eq.\;\eqref{eq:ckstar} is equivalent to 
$\bmm{t}_{k\star} = -i\alpha\left(\bmm{D}_{dd}^{\circ}\right)^{-1}\bmm{Q}\left(\bmm{1} + \alpha\bmm{\Lambda}\right)^{-1}\bmm{Q}^\dagger\left(\bmm{D}^{\circ}_{dd}\right)^{-1} \bmm{d}_k / 2.$
Substituting into Eq.\;\eqref{eq:ckstar} and factoring again, Eq.\;\eqref{eq:alphaSolve} becomes
\begin{equation}
    f\left(\alpha\right) = \frac{\alpha^2}{4}\sum_j \lvert b_{kj}\rvert^2 \frac{2 + \alpha\lambda_j}{\left(1 + \alpha\lambda_j\right)^2},
    \nonumber
\end{equation}
with $\bmm{b}_k = \bmm{Q}^\dagger \left(\bmm{D}^{\circ}_{dd}\right)^{-1}\bmm{d}_k$.
From this form, the required value of $\alpha$ can be determined in a numerically stable manner, i.e. avoiding $\left(\bmm{D}^{\circ}_{dd}\right)^{-1}$, by solving the generalized eigenvalue problem $\bmm{A}\bmm{R} = \left(\bmm{D}^{\circ}_{dd}\right)^2\bmm{R}\bmm{\Lambda}$ and computing $\bmm{b}_k = \bmm{R}^\dagger \bmm{d}_k$.
Since finding $\bmm{\Lambda}$ has no dependence on $\alpha$, the generalized eigenvalue problem must be solved only once to obtain a computationally efficient form of $f(\alpha)$.
\\ \\
A bracketing interval for $f(\alpha)$ can also be found once $\bmm{\Lambda}$ is computed.
Note that, $\forall p_i\in\left\{-\lambda_{j}^{-1} | \lambda_{j} < 0\right\}$, $\lim_{\alpha\to p_i}f(\alpha)\to\infty$. 
Let $p_{\max}$ be the maximum such pole, and $x > p_{\max}\ni f(x) < 0$ (such an $x$ can always be found by taking $x_0 = 2p_\text{max}$ and iterating $x_{n+1} = 2x_n$ until $f(x_n) < 0$).
The form of $f(\alpha)$ in Eq\;\eqref{eq:alphaSolve} makes it clear that $\lim_{\alpha\to\infty}f(\alpha)\to-\infty$.
The interval $\left(p_{\max},\, x_n\right]$ is then a bracketing interval for the last zero of $f(\alpha)$.
Once the value of $\alpha$ is known a solution to Eq.\;\eqref{eq:mainForm} is computed using the above factorization: $\xi_k^2 = \frac{\alpha_{\circ}^2}{4}\sum_j \frac{\lvert b_{kj}\rvert^2}{\left(1 + \alpha_{\circ} \lambda_j\right)^2}$, where $\alpha_{\circ}$ is the last zero of $f$.
\subsection{System Design}
\noindent
With the sender and receiver occupying disjoint volumes, field degrees of freedom are concatenated as $[\bm{s}\ \bm{r}]^\top$ and thus live in $u = s\sqcup r$.
Writing $\bmm{\iota}_{r}: r\hookrightarrow u$ for the inclusion of the receiver in $u$ and $\bmm{\Pi}_{s},\,\bmm{\Pi}_r : u\rightarrow s,\,r$ for the clipping projections of $u$ onto each component, the key operator in Eq.\ \eqref{eq:wChainProg} is applied as $\left(-\bmm{G}^{\circ}_{ru}\right)^{\mathtt{a}} = \left(-\bmm{\iota}_{r} \bmm{G}^{\circ}_{rs}\bmm{\Pi}_{s}\right)^{\mathtt{a}} - \bmm{\iota}_r\left(\bmm{G}^{\circ}_{rr}\right)^{\mathtt{a}}\bmm{\Pi}_r$.
Being Hermitian, its eigendecomposition $\left\{\gamma_{k}, \bm{g}_{k}\right\}$ is computed matrix-free by the randomized scheme of the previous subsection. Eigenvectors with $\gamma_{k}\leq 0$ are discarded, realizing the positive restriction $\left(-\bmm{G}^{\circ}_{ru}\right)^{\mathtt{a}}_{+}$ and fixing channel count as its numerical rank $m$.
\\ \\
The optimization is then projected into the $m$-dimensional span of $\left\{\bm{g}_{k}\right\}$. 
In this basis, the rank-removed objective operator $\bmm{B}_{n} = (4/\zeta) \sum_{k\geq n}\gamma_{k}\,\bm{g}_{k}\otimes\bm{g}_{k}^\dagger$ is diagonal, and the constraint operator $\bmm{C} = \zeta^{-1}\bmm{\Pi}_s + \left(-\bmm{G}^{\circ}_{ru}\right)^{\mathtt{a}}_{+} + \bmm{G}^{\circ \mathtt{a}}_{uu}$ is a dense $m\times m$ Hermitian matrix formed through $m$ operator applications. 
For each channel index $n$, the Lagrangian dual reduces to the generalized eigenvalue problem $\bmm{B}_{n}\bm{v} = \lambda \bmm{C}\bm{v}$. 
Expanding a unit probe vector $\bm{s}_{j}$ in the generalized eigenbasis, we recover a root finding problem very similar to the one discussed above. Probe vectors confined to the sender are constructed by orthonormalizing the sender projections $\bmm{\Pi}_{s}\bm{g}_{k}$ in descending order of $k$ (a reverse Gram-Schmidt procedure), so that the probe associated with the index $j$ is orthogonal to the sender content of all higher-index eigenvectors. 
This guarantees orthogonality between the $k^{th}$ test probe and the sender content of $\left\{\bmm{g}_{j}\right\}_{j>k}$.
\\ \\
The reported bound on $\sigma_{n}\left(\bmm{P}_{rs}\right)$ is the minimum of three quantities: the numerical dual above, and analytic passivity bounds from Eq.\ \eqref{eq:anaResultPrsAmp}'s functional form using either $\kappa_{n} = \zeta^2\sigma^2_{n}\left(\bmm{G}^{\circ}_{rs}\right)$ or $\kappa_{n} = \zeta\max\left(\sigma_{n}\left(\left(-\bmm{G}^{\circ}_{ru}\right)^{\mathtt{a}}\right), 0\right)$.
\subsection{Two-sphere Analytics}
\noindent
The analytic comparison of Fig.\;\ref{fig:traceBounds} (d) for two identical homogeneous silica \cite{palik1998handbook} balls follows the vector-spherical-wave (VSW) multiple-scattering formulation of Narayanaswamy and Chen. 
In this subsection, terminology and notation pertaining to the VSW formalism follows directly from this reference, Ref.\;\cite{narayanaswamy2008thermal}.
\\ \\
Axial translation coefficients are evaluated in closed Wigner-3$j$ form and validated by verifying agreement with a formula-free projection constructed from the VSW orthonormality condition under coordinate translation. 
Spherical radial functions at high order and small separations (where numerical overflow and underflow occurs in double precision floating point arithmetic) are computed in extended precision by Miller's downward recurrence for spherical Bessel functions of the first kind and upward recurrence for the second kind. Final calculations for $\Phi$ and other related quantities are in agreement with the reference paper.
\subsection{Numerical Fidelity of the Bounds}
\noindent
Due to the inherently randomized nature of the scheme for computing the ordered singular value bounds in three dimensions, the main lessons that should be drawn from Fig.\;2 are comparative: it is better to focus on index trends (decay rates, relative spacing, and cross-scenario ordering) of $\bmm{G}_{rs}$ over the absolute magnitudes of any given singular value. 
Although we firmly believe that the current bounds are quantitatively accurate, variations within an order of unity would not be completely surprising. 
\bibliography{esaLibF}
\end{document}